\documentclass[
aps,
twocolumn,
10pt,
prd,
floatfix
]{revtex4-2}

\usepackage{overpic}
\usepackage{graphicx}
\usepackage{epsfig}
\usepackage{dcolumn}
\usepackage{bm}
\usepackage{rotating}
\usepackage{multirow}
\usepackage{booktabs}
\usepackage{appendix}
\usepackage{subfigure}
\usepackage{hhline}
\usepackage{amssymb}
\usepackage{lineno}
\usepackage{hyperref}
\usepackage{mathrsfs}
\usepackage{verbatim}
\usepackage{amsmath}
\usepackage{amssymb}
\usepackage{amsfonts}
\usepackage{amsbsy}
\usepackage{float}

\usepackage{CJKutf8}
\usepackage{natbib}
\usepackage{verbatim}
\usepackage{lineno}
\usepackage{hyperref}
\hypersetup{
    hidelinks,
    colorlinks=true,
    allcolors=blue,
    pdfstartview=Fit,
    breaklinks=true
}

\include{def-com}

\let\oldequation\equation
\let\oldendequation\endequation

\renewenvironment{equation}
  {\linenomathNonumbers\oldequation}
  {\oldendequation\endlinenomath}

\begin{document}

\title{\boldmath Search for $\eta_{1}(1855)$ in $\chi_{cJ}\to\eta\eta\eta^{\prime}$ decays}

\author{
\begin{small}
  \begin{center}
M.~Ablikim$^{1}$, M.~N.~Achasov$^{4,c}$, P.~Adlarson$^{77}$, X.~C.~Ai$^{82}$, R.~Aliberti$^{36}$, A.~Amoroso$^{76A,76C}$, Q.~An$^{73,59,a}$, Y.~Bai$^{58}$, O.~Bakina$^{37}$, Y.~Ban$^{47,h}$
% ...
\\
\vspace{0.2cm}
(BESIII Collaboration)\\
\vspace{0.2cm} {\it
$^{1}$ Institute of High Energy Physics, Beijing 100049, People's Republic of China\\
% ...
$^{4}$ Budker Institute of Nuclear Physics SB RAS (BINP), Novosibirsk 630090, Russia\\
% ...
$^{36}$ Johannes Gutenberg University of Mainz, Johann-Joachim-Becher-Weg 45, D-55099 Mainz, Germany\\
$^{37}$ Joint Institute for Nuclear Research, 141980 Dubna, Moscow region, Russia\\
% ...
$^{47}$ Peking University, Beijing 100871, People's Republic of China\\
% ...
$^{58}$ Southeast University, Nanjing 211100, People's Republic of China\\
$^{59}$ State Key Laboratory of Particle Detection and Electronics, Beijing 100049, Hefei 230026, People's Republic of China\\
% ...
$^{73}$ University of Science and Technology of China, Hefei 230026, People's Republic of China\\
% ...
$^{76}$ University of Turin and INFN, (A)University of Turin, I-10125, Turin, Italy; (B)University of Eastern Piedmont, I-15121, Alessandria, Italy; (C)INFN, I-10125, Turin, Italy\\
$^{77}$ Uppsala University, Box 516, SE-75120 Uppsala, Sweden\\
% ...
$^{82}$ Zhengzhou University, Zhengzhou 450001, People's Republic of China\\

\vspace{0.2cm}
$^{a}$ Deceased\\
% ...
$^{c}$ Also at the Novosibirsk State University, Novosibirsk, 630090, Russia\\
% ...
$^{h}$ Also at State Key Laboratory of Nuclear Physics and Technology, Peking University, Beijing 100871, People's Republic of China\\
% ...

}

\end{center}
\end{small}
}


\begin{abstract}
Based on a sample of $2.7\times10^{9}$ $\psi(3686)$ events collected by the BESIII detector operating at the BEPCII collider, an analysis of the decay $\psi(3686)\to\gamma\chi_{cJ}, \chi_{cJ}\to\eta\eta\eta^{\prime}$ is performed. The decay modes $\chi_{c1}$ and $\chi_{c2}\to\eta\eta\eta^{\prime}$ are observed for the first time, and their corresponding branching fractions are determined to be $\mathcal{B}(\chi_{c1}\to\eta\eta\eta^{\prime}) = (1.40\, \pm 0.13\, (\text{stat.}) \pm 0.09\, (\text{sys.})) \times 10^{-4}$ and $\mathcal{B}(\chi_{c2}\to\eta\eta\eta^{\prime}) = (4.18\, \pm 0.84\, (\text{stat.}) \pm 0.48\, (\text{sys.})) \times 10^{-5}$. An upper limit on the branching fraction of $\chi_{c0}\to\eta\eta\eta^{\prime}$ is set as $2.59 \times 10^{-5}$ at 90\% confidence level (CL). A partial wave analysis (PWA) of the decay $\chi_{c1}\to\eta\eta\eta^{\prime}$ is performed to search for the $1^{-+}$ exotic state $\eta_1(1855)$. The PWA result indicates that the structure in the $\eta\eta^{\prime}$ mass spectrum is mainly attributed to the $f_0(1500)$, while in the $\eta\eta$ mass spectrum, it is primarily the $0^{++}$ phase space. The upper limit of $\mathcal{B}(\chi_{c1}\to\eta_{1}(1855)\eta) \cdot \mathcal{B}(\eta_{1}(1855)\to\eta\eta^{\prime})< 9.79 \times 10^{-5}$ is set based on the PWA at 90\% CL.

\end{abstract}

\maketitle
\oddsidemargin -0.2cm
\evensidemargin -0.2cm

\section{\boldmath Introduction}
The quark model describes mesons as composed of a quark and an antiquark ($q\bar{q}$), while quantum chromodynamics (QCD) allows the existence of exotic states. Exploring spin-exotic states is an important area of research. These states have exotic quantum numbers beyond those in the quark model, such as $J^{PC}=0^{--}$, even$^{+-}$, and odd$^{-+}$, which can be easily identified.

A hybrid is an exotic state composed of a quark, an antiquark and an excited gluon field, which reflects the gluonic degrees of freedom.
The $1^{-+}$ hybrid state, predicted by lattice QCD (LQCD) as the lightest hadron with spin-exotic quantum numbers, is estimated to have a mass in \text{1.7-2.1 GeV/$c^2$} \cite{Dudek:2013yja}.
LQCD has recently investigated the production of $1^{-+}$ hybrid in $J/\psi$ radiative decay and its possible decay modes \cite{Chen:2022isv, Liang:2024lon}.
The isovector state $\pi_1(1600)$ has long been considered an experimental candidate for the $1^{-+}$ spin-exotic nonet \cite{Meyer:2015eta, Huang:2010dc, Chen:2010ic, Eshraim:2020ucw}. 
CLEO-c has found a clear P-wave signal in the $\eta^{\prime}\pi^{\pm}$ system around 1.6 GeV/$c^2$ from the $\chi_{c1}\to\eta^{\prime}\pi^+\pi^-$ process \cite{CLEO:2011upl}.
Recently, the $1^{-+}$ isoscalar state $\eta_1(1855)$ was discovered in the $\eta\eta^{\prime}$ system by BESIII via $J/\psi\to\gamma\eta\eta^{\prime}$ \cite{BESIII:2022riz, BESIII:2022iwi} and it is considered to be a partner of $\pi_1(1600)$, marking a new research direction for spin-exotic states. 
These research results help in understanding the production mechanism of the $1^{-+}$ spin-exotic state.
It is possible to observe $\eta_1(1855)$ via $\chi_{c1}\to\eta_1(1855)\eta$, with $\eta_1(1855)\to\eta\eta^{\prime}$.

In this paper, based on a sample of $2.7\times10^{9}$ $\psi(3686)$ events collected by the BESIII detector \cite{BESIII:2009fln}, an analysis of the decay $\psi(3686)\to\gamma\chi_{cJ}$, $\chi_{cJ}\to\eta\eta\eta^{\prime}$ is performed to search for $\eta_{1}(1855)$.
The $\eta^{\prime}$ is reconstructed via $\eta^{\prime}\to\gamma\pi^+\pi^-$ and $\eta^{\prime}\to\eta\pi^+\pi^-$, and the $\eta$ is reconstructed via $\gamma\gamma$.

\section{\boldmath BESIII Detector and Monte Carlo Simulation}

The BESIII detector~\cite{BESIII:2009fln} records symmetric $e^+e^-$ collisions provided by the BEPCII storage ring~\cite{Yu:2016cof}, with a center-of-mass energy ranging from 1.84 to 4.95~GeV and a peak luminosity of $1.1\times10^{33}$~cm$^{-2}$s$^{-1}$ achieved at $\sqrt{s}=3.773$ GeV. BESIII has collected large data samples in this energy region~\cite{BESIII:2020nme}. The cylindrical core of the BESIII detector covers 93\% of the full solid angle and consists of a helium-based multilayer drift chamber~(MDC), a time-of-flight system~(TOF), and a CsI~(Tl) electromagnetic calorimeter~(EMC), which are all enclosed in a superconducting solenoidal magnet providing a 1.0~T magnetic field. The solenoid is supported by an octagonal flux-return yoke with resistive plate counter muon identification modules interleaved with steel. The charged-particle momentum resolution at $1~{\rm GeV}/c$ is $0.5\%$, and the resolution of the specific ionization energy (d$E$/d$x$) is $6\%$ for electrons from Bhabha scattering. The EMC measures photon energies with a resolution of $2.5\%$ ($5\%$) at $1$~GeV in the barrel (end-cap) region. The time resolution in the TOF plastic scintillator barrel region is 68~ps, while that in the end-cap region was 110~ps. The end-cap TOF system was upgraded in 2015 using multi-gap resistive plate chamber technology, providing a time resolution of 60~ps, which benefits 85\% of the data used in this analysis~\cite{Li:2017jpg, Guo:2017sjt, Cao:2020ibk}.

Monte Carlo (MC) simulated samples, produced with  {\sc geant4}-based~\cite{GEANT4:2002zbu} software, which includes the geometric description of the BESIII detector and the detector response, are used to determine detection efficiencies and to estimate backgrounds. The simulation models the beam energy spread and initial state radiation (ISR) in the $e^+e^-$ annihilations with the generator {\sc kkmc}~\cite{Jadach:1999vf, Jadach:2000ir}. The inclusive MC sample includes the production of the $\psi(3686)$ resonance, the ISR production of the $J/\psi$, and the continuum processes incorporated in {\sc kkmc}~\cite{Jadach:1999vf, Jadach:2000ir}. Known decay modes are modeled with {\sc evtgen}~\cite{Lange:2001uf, Ping:2008zz} using branching fractions taken from the Particle Data Group (PDG)~\cite{ParticleDataGroup:2024cfk}. The remaining unknown charmonium decays are modeled with {\sc lundcharm}~\cite{Chen:2000tv, Yang:2014vra}. Final state radiation~(FSR) from charged final state particles is incorporated using {\sc photos} package~\cite{Barberio:1990ms}. Exclusive MC samples are generated to determine the detection efficiency and optimize selection criteria.
The signal is simulated with the E1 transition $\psi(3686)\to\gamma\chi_{cJ}$, where the polar angle $\theta$ of the radiative photon in the $e^+e^-$ center-of-mass frame is distributed according to $(1+\lambda cos^{2}\theta)$. For $J=0$, 1 and 2, $\lambda$ is set to be 1, $-\frac{1}{3}$ and $\frac{1}{13}$, respectively \cite{Ping:2008zz, Lange:2001uf}. 
The decay $\eta^{\prime}\to\gamma\pi^{+}\pi^{-}$ is simulated by a model that considers both the $\rho-\omega$ interference and the box anomaly \cite{BESIII:2017kyd}, while the decay $\eta^{\prime}\to\eta\pi^{+}\pi^{-}$ is simulated with a model based on the measured matrix elements \cite{BESIII:2017djm}. Other processes are generated uniformly in phase space (PHSP).

\section{\boldmath Event Selection}
\label{sec:selection}

Charged tracks detected in the MDC are required to
have a polar angle ($\theta$) within the range of $|\cos\theta|<0.93$, where $\theta$ is defined with respect to the $z$-axis, the symmetry axis of the MDC. The distance of the closest approach to the interaction point (IP) must be less than 10~cm along the $z$-axis, and less than 1~cm in the transverse plane. Two good charged tracks are required in the final state, and the total charge must be equal to zero.

The particle identification (PID) for charged tracks combines measurements of the d$E$/d$x$ in the MDC and the flight time in the TOF to form likelihoods $\mathcal{L}(h)$ $(h=p, K, \pi)$ for each hadron ($h$) hypothesis. A charged track is identified as a pion when the pion hypothesis yields the maximum likelihood value, i.e. $\mathcal{L}(\pi)>\mathcal{L}(K)$, and $\mathcal{L}(\pi)>\mathcal{L}(p)$. The two good charged tracks must both be identified as pions.

In the selection of good photon candidates, the deposited energy for a cluster is required to be larger than 25~MeV in the barrel ($ |\cos \theta| < 0.80 $) or 50~MeV in the end-cap ($ 0.86 < |\cos \theta| < 0.92$) regions.   To exclude spurious photons caused by hadronic interactions and final state radiation, photon candidates must be at least $10^{\circ}$ away from any charged tracks when extrapolated to the EMC.  To suppress electronic noise and unrelated showers,  the difference between the EMC time and the event start time is required to be within [0, 700]~ns.

\begin{figure*}[!htbp]
    \centering
    \subfigure[\label{fig_Metap_gampipi}]{\includegraphics[width=0.46\textwidth]{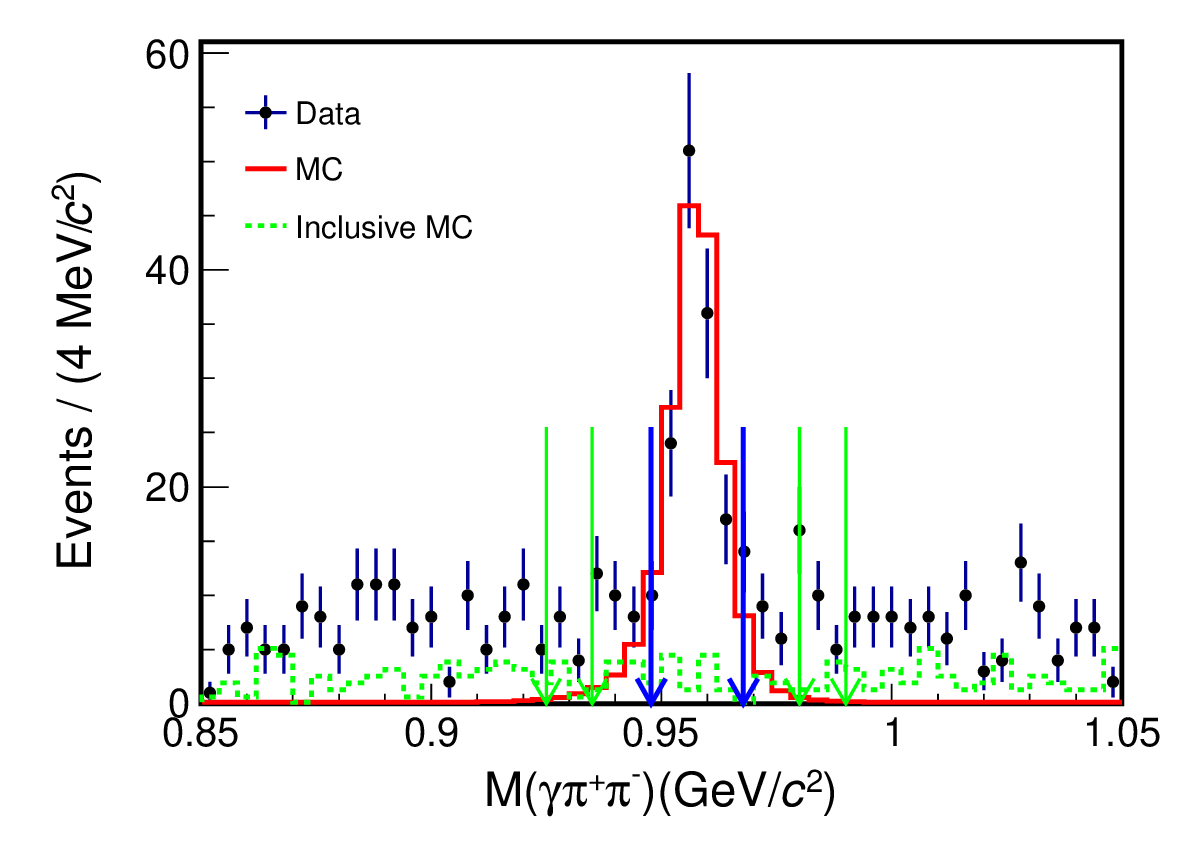}}
    \subfigure[\label{fig_Metap_etapipi}]{\includegraphics[width=0.46\textwidth]{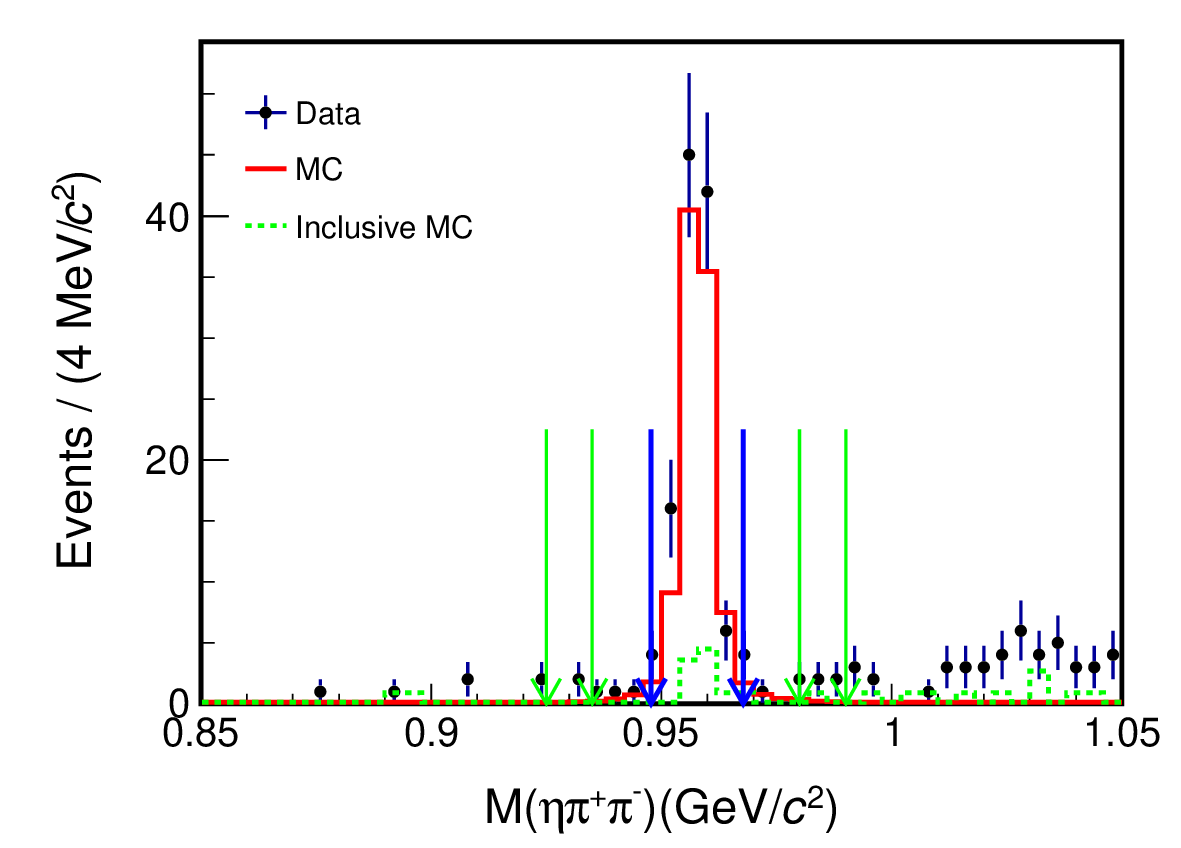}}
    \caption{Invariant mass distributions of (a) $\eta^{\prime}\to\gamma\pi^+\pi^-$ and (b) $\eta^{\prime}\to\eta\pi^+\pi^-$ within all $\chi_{cJ}$ range (3.34-3.64 GeV/$c^{2}$). The black points with error bars are data, the red solid histograms are the signal MC and the green dashed histograms are the inclusive MC. The events between the blue arrows are the selected signal events, and the events between the green arrows on the same side are selected sideband events.}
\end{figure*}

For the $\psi(3686)\to\gamma\chi_{cJ}$, $\chi_{cJ}\to\eta\eta\eta^{\prime}$, $\eta^{\prime}\to\gamma\pi^+\pi^-$ mode, events are reconstructed with two oppositely charged tracks and at least six candidate photons.
A six-constraint (6C) kinematic fit under the hypothesis of $\psi(3686)\to\gamma\gamma\eta\eta\pi^+\pi^-$ is performed by imposing the  energy-momentum conservation and constraining the mass of each pair of photons to the nominal $\eta$ mass from the PDG \cite{ParticleDataGroup:2024cfk}.
If there is more than one combination, the one with the minimum $\chi^2$ value of the fit (denoted as $\chi^2_{6\rm{C}}$) is retained.
The result $\chi^2_{6\rm{C}}$ is required to be less than 20, which is obtained by optimizing the figure of merit (FOM). The FOM is defined as $\text{FOM}=\frac{S}{\sqrt{S+B}}$, where $S$ is the normalized number of signal events in the signal MC samples and $(S+B)$ corresponds to the number of data events.
To suppress backgrounds with five or seven photons in the final state, 4C kinematic fits are performed by imposing the energy-momentum conservation under the hypotheses of $\psi(3686)\to 5\gamma\pi^{+}\pi^{-}$, $\psi(3686)\to 6\gamma\pi^{+}\pi^{-}$ and $\psi(3686)\to 7\gamma\pi^{+}\pi^{-}$.
The $\chi^{2}_{\rm{4C}}(6\gamma\pi^{+}\pi^{-})$ is required to be less than all possible $\chi^{2}_{\rm{4C}}(5\gamma\pi^{+}\pi^{-})$ and $\chi^{2}_{\rm{4C}}(7\gamma\pi^{+}\pi^{-})$.
To reconstruct the $\eta^{\prime}$ candidate, the $\gamma\pi^+\pi^-$ combination with the minimum $|M(\gamma\pi^{+}\pi^{-})-m_{\eta^{\prime}}|$ is selected, where $m_{\eta^{\prime}}$ is the nominal $\eta^{\prime}$ mass taken from the PDG \cite{ParticleDataGroup:2024cfk}.
Events with $|M(\gamma\pi^{+}\pi^{-})-m_{\eta^{\prime}}|<0.01$ GeV/$c^{2}$ are selected for further analysis.
To suppress backgrounds containing $\pi^0$, events with $|M(\gamma\gamma)-m_{\pi^{0}}|<0.02$ GeV/$c^{2}$ are rejected, where $M(\gamma\gamma)$ is the invariant mass of each photon pair except the two-photon pairs assigned to the $\eta$, and $m_{\pi^{0}}$ is the nominal $\pi^{0}$ mass \cite{ParticleDataGroup:2024cfk}.
The invariant mass distribution of $\gamma\pi^+\pi^-$ is shown in \text{Fig.  \ref{fig_Metap_gampipi}}.

For the decays $\psi(3686)\to\gamma\chi_{cJ}$, $\chi_{cJ}\to\eta\eta\eta^{\prime}$, $\eta^{\prime}\to\eta\pi^+\pi^-$, events are reconstructed with two oppositely charged tracks and at least seven candidate photons.
A seven-constraint (7C) kinematic fit under the hypothesis of $\psi(3686)\to\gamma\eta\eta\eta\pi^+\pi^-$ is performed by imposing the  energy-momentum conservation and constraining the mass of each pair of photons to $m_{\eta}$.
If there is more than one combination, the one with the minimum $\chi^2$ value of the fit (denoted as $\chi^2_{7\rm{C}}$) is retained.
The result $\chi^2_{7\rm{C}}$ is required to be less than 55, which is obtained by optimizing the FOM,
similarly to the optimization performed for the $\eta^{\prime}\to\gamma\pi^+\pi^-$ mode.
To suppress backgrounds with eight photons in the final state, 4C kinematic fits are performed by imposing the energy-momentum conservation under the hypotheses of $\psi(3686)\to 6\gamma\pi^{+}\pi^{-}$, $\psi(3686)\to 7\gamma\pi^{+}\pi^{-}$ and $\psi(3686)\to 8\gamma\pi^{+}\pi^{-}$.
$\chi^{2}_{\rm{4C}}(7\gamma\pi^{+}\pi^{-})$ is required to be less than $\chi^{2}_{\rm{4C}}(8\gamma\pi^{+}\pi^{-})$.
All the selected events satisfy $\chi^{2}_{\rm{4C}}(7\gamma\pi^{+}\pi^{-})<\chi^{2}_{\rm{4C}}(6\gamma\pi^{+}\pi^{-})$. 
The $\eta\pi^+\pi^-$ combination with the minimum $|M(\eta\pi^{+}\pi^{-})-m_{\eta^{\prime}}|$ is used to reconstruct the $\eta^{\prime}$ candidate.
Events with $|M(\eta\pi^{+}\pi^{-})-m_{\eta^{\prime}}|<0.01$ GeV/$c^{2}$ are selected for further analysis.
Backgrounds containing $\pi^0$ are suppressed by rejecting events with $|M(\gamma\gamma)-m_{\pi^{0}}|<0.02$ GeV/$c^{2}$.
The invariant mass distribution of $\eta\pi^+\pi^-$ is shown in \text{Fig. \ref{fig_Metap_etapipi}}.

\section{\boldmath Measurement of $\chi_{cJ}\to\eta\eta\eta^{\prime}$ }

\subsection{\boldmath Background estimation}

Potential backgrounds are studied using the inclusive MC samples of $\psi(3686)$ decays.
The backgrounds are found to fall into three categories: non-$\chi_{cJ}$, non-$\eta^{\prime}$, and processes involving both $\eta^{\prime}$ and $\chi_{cJ}$.
Non-$\chi_{cJ}$ backgrounds are modeled by a polynomial function when fitting the invariant mass distributions of $\eta\eta\eta^{\prime}$.
Non-$\eta^{\prime}$ backgrounds are estimated using the $\eta^{\prime}$ mass sidebands in the data. The sideband regions for $\eta^{\prime}$ are defined as [0.925, 0.935] and [0.98, 0.99] GeV/$c^{2}$.
The $\eta^{\prime}$ signal shape is determined from the shape of the signal MC sample, and the $\eta^{\prime}$ sideband shape is described by a linear function.
The backgrounds containing both $\chi_{cJ}$ and $\eta^{\prime}$ are peaking backgrounds.
To estimate the peaking backgrounds, exclusive MC samples with high statistics are generated for those peaking background processes as indicated in the inclusive MC.
The peaking background contributions are normalized with the corresponding branching fractions from PDG and the efficiency obtained from the background MC samples, which are then fixed in the fit.
The peaking backgrounds for the decay mode $\eta^{\prime}\to\gamma\pi^+\pi^-$ primarily arise from $\chi_{c1}\to\gamma J/\psi, J/\psi\to\gamma\eta\eta^{\prime}$ and $\chi_{c2}\to\gamma J/\psi, J/\psi\to\gamma\eta\eta^{\prime}$, with estimated yields (fractions) of 2.35 (1.7\%) and 1.14 (0.8\%), respectively.
Similarly, the peaking backgrounds for the decay mode $\eta^{\prime}\to\eta\pi^+\pi^-$ are mainly from $\chi_{c1}\to\gamma J/\psi, J/\psi\to\gamma\eta\eta’$, $\chi_{c2}\to\gamma J/\psi, J/\psi\to\gamma\eta\eta’$ and $\chi_{c0}\to\eta^{\prime}\eta^{\prime}$, with estimated yields (fractions) of 1.97 (1.7\%), 0.82 (0.7\%) and 1.25 (1.1\%), respectively.
The backgrounds with a fraction of no more than 1.0\% are neglected.

\subsection{\boldmath Signal extraction}
\begin{figure*}[!htbp]
    \centering
    \subfigure[]{\includegraphics[width=0.46\textwidth]{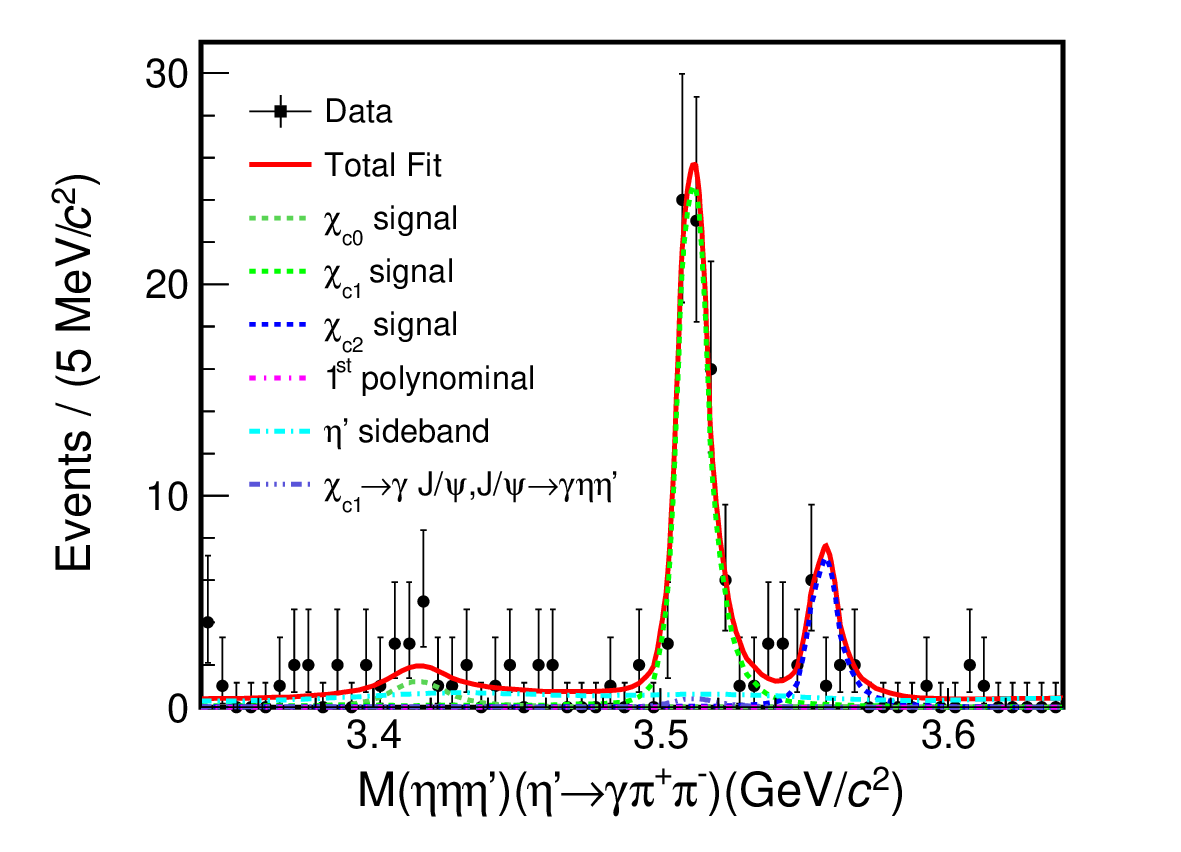}}
    \subfigure[]{\includegraphics[width=0.46\textwidth]{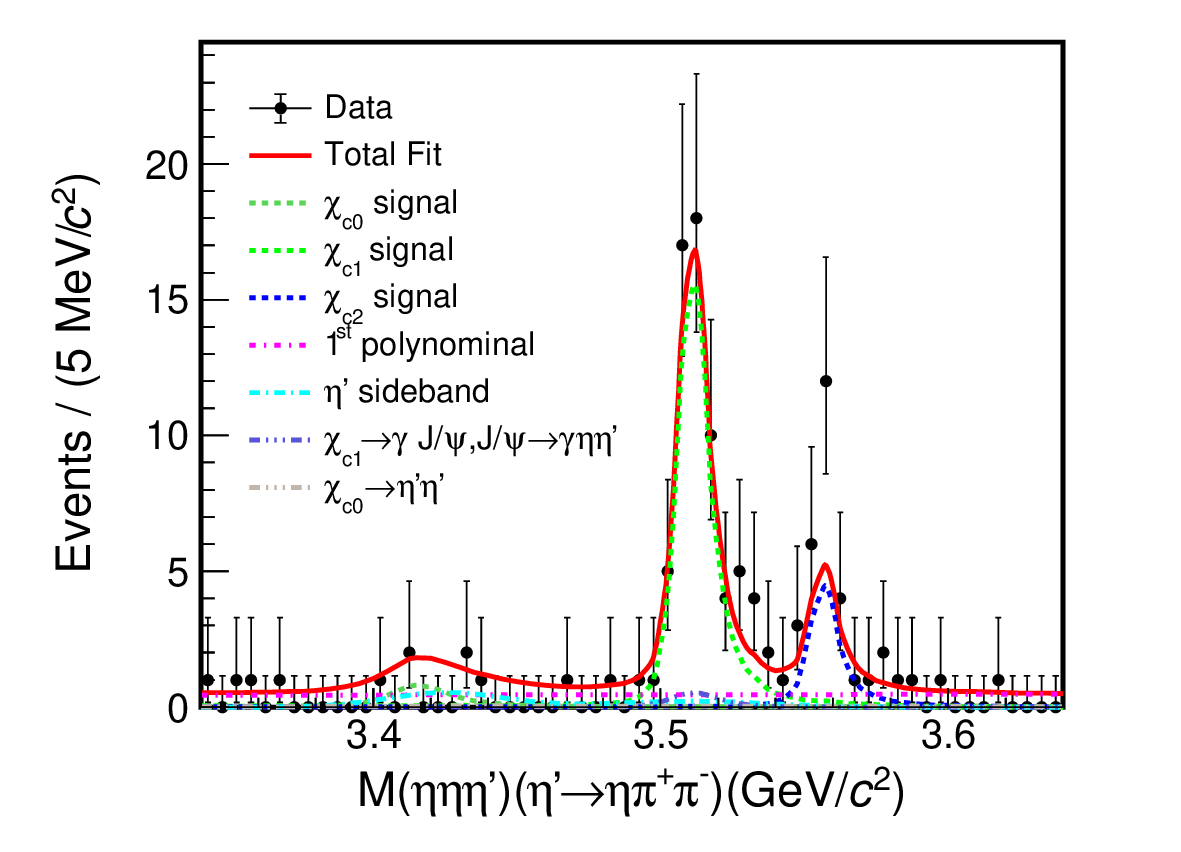}}
    \caption{The simultaneous fit results for the invariant mass distributions of $\eta\eta\eta^{\prime}$ in two decay modes of $\eta^{\prime}$ (a) $\eta^{\prime}\to\gamma\pi^+\pi^-$ and (b) $\eta^{\prime}\to\eta\pi^+\pi^-$. The dots with error bars represent the data, the red solid curves show the fit results, and the dark green, light green, and blue dashed lines represent the signals of $\chi_{c0}$, $\chi_{c1}$ and $\chi_{c2}$, while the other dashed lines represent the backgrounds.}
    \label{fig_fitMetaetaetap}
\end{figure*}

The $\eta\eta\eta^{\prime}$ invariant mass distributions for the $\eta^{\prime}\to\gamma\pi^{+}\pi^{-}$ and $\eta^{\prime}\to\eta\pi^{+}\pi^{-}$ modes are shown in \text{Fig. \ref{fig_fitMetaetaetap}}.
We perform a simultaneous unbinned maximum likelihood fit to the $\eta\eta\eta^{\prime}$ distributions in the range of [3.34, 3.64] GeV/$c^{2}$.
The $\chi_{c0}$, $\chi_{c1}$, and $\chi_{c2}$ signals are modeled using the shape of MC samples convolved with a Gaussian function to account for the mass resolution and describe the difference between data and MC simulation.
Parameters of the Gaussian function are shared among the $\chi_{c0}$, $\chi_{c1}$, and $\chi_{c2}$ signals and left free in the fit.
The peaking backgrounds are normalized and fixed based on the MC simulations.
The non-$\eta^{\prime}$ background events are described by the $\eta^{\prime}$ mass sidebands and are fixed in the fit.
The remaining background is described by a linear function with free parameters.
In the simultaneous fit, the fitted yields of $\mathcal{N}_{\text{sig}, 1}$ and $\mathcal{N}_{\text{sig}, 2}$ are constrained to the common number of produced $\chi_{cJ}\to\eta\eta\eta^{\prime}$ events $\mathcal{N}^{\chi_{cJ}}_{\text{sig}}$, which is defined as:

\begin{equation}
    \label{eq_Nobs}
   \begin{split}
    &\mathcal{N}^{\chi_{cJ}}_{\text{sig}}=\frac{\mathcal{N}_{\text{sig}, 1}}{\epsilon_{1} \cdot \mathcal{B}(\eta^{\prime}\to\gamma\pi^{+}\pi^{-}) \cdot \mathcal{B}(\eta\to\gamma\gamma)^2}  \\
    &= \frac{\mathcal{N}_{\text{sig}, 2}}{\epsilon_{2} \cdot \mathcal{B}(\eta^{\prime}\to\eta\pi^{+}\pi^{-}) \cdot \mathcal{B}(\eta\to\gamma\gamma)^3},
   \end{split}
\end{equation}
where indices 1 and 2 represent $\eta^{\prime}\to\gamma\pi^{+}\pi^{-}$ and $\eta^{\prime}\to\eta\pi^{+}\pi^{-}$, respectively.
$\mathcal{N}_{\text{sig}, 1}$ and $\mathcal{N}_{\text{sig}, 2}$ are the numbers of observed $\chi_{cJ}$ events, $\epsilon_{1}$ and $\epsilon_{2}$ are the corresponding detection efficiencies, and the branching fractions are obtained from the PDG \cite{ParticleDataGroup:2024cfk}.
The fit result is shown in \text{Fig. \ref{fig_fitMetaetaetap}}.

The branching fractions of $\chi_{cJ}\to\eta\eta\eta^{\prime}$ are calculated by
\begin{equation}
\label{eq_chicj_to_etaetaetap}
\mathcal{B}(\chi_{cJ}\to\eta\eta\eta^{\prime}) = \frac{ \mathcal{N}^{ \chi_{cJ}}_{\text{sig}} }{ N_{\psi(3686)} \cdot \mathcal{B}(\psi(3686)\to\gamma\chi_{cJ})},
\end{equation}
where $\mathcal{N}^{ \chi_{cJ}}_{\text{sig}}$ is defined in Eq. \ref{eq_Nobs},
$N_{\psi(3686)}$ is the number of $\psi(3686)$ events and $\mathcal{B}(\psi(3686)\to\gamma\chi_{cJ})$ is the branching fraction of $\psi(3686)\to\gamma\chi_{cJ}$ obtained from PDG \cite{ParticleDataGroup:2024cfk}.
Since no clear signal for $\chi_{c0}$ exists, a Bayesian method is used to obtain the upper limit of the signal yield at 90\% confidence level (CL).
To determine the upper limit of the signal yield, the distribution of normalized likelihood values for a series of expected signal yields is taken as the probability density function (PDF).
In obtaining the distribution of likelihood values, the signals of $\chi_{c1}$ and $\chi_{c2}$ are described by the MC shape function convolved with a Gaussian function, and their parameters are free.
The upper limit of the signal yields at the 90\% CL, denoted as $N^{\text{UL}}$, is defined as the number of events where 90\% of the PDF area lies above zero.
These results of branching fractions are listed in \text{Table \ref{tab_branchresult}}.
The statistical significances of $\chi_{c0}$, $\chi_{c1}$ and $\chi_{c2}$ are evaluated based on the change of the likelihood value when they are included, taking into account the change in the number of degrees of freedom.

\begin{table*}[!htbp]
	\setlength{\abovecaptionskip}{0cm}
	\setlength{\belowcaptionskip}{0.5cm}
    \small
    \centering
    \caption{
    The numbers of the observed $\chi_{cJ}$ events ($\mathcal{N}_{\text{sig},1}$ and $\mathcal{N}_{\text{sig},2}$), 
    corresponding detection efficiencies ($\epsilon_1$ and $\epsilon_2$), 
    the number of produced $\chi_{cJ}\to\eta\eta\eta^{\prime}$ events ($\mathcal{N}^{ \chi_{cJ}}_{\text{sig}}$,  which is defined in Eq. \ref{eq_Nobs}), 
    branching fractions ($\mathcal{B}$), and the statistical significance, respectively. Indices 1 and 2 represent $\eta^{\prime}\to\gamma\pi^+\pi^-$ and $\eta^{\prime}\to\eta\pi^+\pi^-$, respectively. The first and second uncertainties are statistical and systematic, respectively.}
    \label{tab_branchresult}
    \begin{tabular}{c|c|c|c|c|c|c|c}
    \hline
        \hline
        $\chi_{cJ}$ & $\mathcal{N}_{\text{sig}, 1}$ & $\epsilon_1$(\%) &  $\mathcal{N}_{\text{sig}, 2}$& $\epsilon_2$(\%) & $\mathcal{N}^{ \chi_{cJ}}_{\text{sig}}$ & $\mathcal{B}(\chi_{cJ}\to\eta\eta\eta^{\prime})$ &  Significance \\
        \hline
$\chi_{c0}$&  $9.1 \pm 3.9$ & $5.23\pm0.02$ &$ 3.8 \pm 1.6$&  $3.86\pm 0.02$ & $3809.6\pm 1607.6$    & - &  $1.4\sigma$ \\
$\chi_{c1}$&  $97.7\pm 9.1$   & $5.76\pm 0.02$ &$38.8 \pm 3.6$  & $4.03\pm 0.02$ & $37016.9\pm  3431.3$& $(1.40 \pm 0.13 \pm 0.09)\times 10^{-4}$ &  $16.8\sigma$ \\
$\chi_{c2}$ &  $26.3\pm 5.3$  & $5.32\pm 0.02$ &$ 10.4 \pm 2.1$ & $3.70\pm 0.02$ & $10783.4\pm  2162.44$ & $(4.18 \pm 0.84 \pm 0.48)\times 10^{-5}$ &  $7.2\sigma$ \\
        \hline
        \hline
    \end{tabular}
\end{table*}

\subsection{Systematic uncertainties}
The systematic uncertainties of $\mathcal{B}(\chi_{cJ}\to\eta\eta\eta^{\prime})$ are summarized below.

\begin{enumerate}
    \item Number of $\psi(3686)$ events. The number of $\psi(3686)$ events is determined by analyzing inclusive hadronic $\psi(3686)$ decays with an uncertainty of 0.5\% \cite{BESIII:2024lks}.

    \item MDC tracking. The data-MC efficiency difference for pion track-finding is studied using a control sample of $J/\psi\to\rho\pi$ and $J/\psi\to p\bar{p}\pi^{+}\pi^{-}$ \cite{BESIII:2013qmu}. The systematic uncertainty from MDC tracking  is 1.0\% per charged pion.

    \item PID. The pion PID efficiency for data agrees within 1.0\% with that of the MC simulation in the pion momentum region, as reported in \cite{BESIII:2013qmu}.  Thus, a systematic uncertainty of 2.0\% is assigned to PID.

    \item Photon reconstruction. For photons detected by the EMC, the detection efficiency is studied using a control sample of $e^+e^-\to\gamma_{\text{ISR}}\mu^+\mu^-$, where ISR stands for initial state radiation. The systematic uncertainty, defined as the relative difference in efficiencies between data and MC simulation, is observed to be up to 0.5\% per photon in both the barrel and end-cap regions.

    \item Intermediate decay branching fractions. The uncertainties on the intermediate decay branching fractions  are taken from the PDG values \cite{ParticleDataGroup:2024cfk}.

    \item Kinematic fit. To investigate the systematic uncertainty associated with the kinematic fit, the track helix parameter correction method is used \cite{BESIII:2012mpj}. The difference in the detection efficiencies with and without the helix correction is taken as the systematic uncertainty.

    \item Mass windows of $\eta^{\prime}$ and $\chi_{cJ}$. The systematic uncertainty from the requirement on the $\eta^{\prime}$ or $\chi_{cJ}$ signal region is estimated by smearing the $\gamma\pi^{+}\pi^{-}$($\eta\pi^{+}\pi^{-}$) and $\eta\eta\eta^{\prime}$ invariant masses in the signal MC sample with a Gaussian function to account for the resolution difference between data and MC simulation. The smearing parameters are determined by fitting the $\gamma\pi^{+}\pi^{-}$($\eta\pi^{+}\pi^{-}$) and $\eta\eta\eta^{\prime}$ invariant mass distributions in data with the MC shape convolved with a Gaussian function. The difference in the detection efficiency, as determined from the signal MC sample with and without the extra smearing, is taken as the systematic uncertainty.

    \item Veto of $\pi^0$.
    Possible systematic effects due to the veto on $\pi^0$ are investigated using the Barlow test \cite{Barlow:2002yb}, by varying the veto criteria between 10 and 50 MeV/$c^2$. The uncertainty is assigned as the difference between the nominal result and the result with $|M(\gamma\gamma)-m_{\pi^0}|<21$ MeV/$c^2$ when 95\% of the candidate events are retained. Since there is no signal in the $\chi_{c0}$ process, we set the value of this systematic uncertainty to be the same as that for $\chi_{c1,\, c2}$.

    \item Fit range. The systematic uncertainties due to different fit ranges are examined by enlarging and shrinking the fit range eight times, with a step of 20 MeV/$c^2$. For each case, the deviation between the alternative and nominal fits is defined as $\xi=\frac{|a_1-a_2|}{\sqrt{|\sigma_1^2-\sigma_2^2|}}$, where $a_1\pm \sigma_1$ and $a_2\pm\sigma_2$ are the nominal fit results and alternative fit results, respectively.  If $\xi$ is less than 2.0, the associated systematic uncertainty is negligible according to the Barlow test \cite{Barlow:2002yb}. The largest difference is assigned as the systematic uncertainty.

    \item Non-peaking background shape.  The systematic uncertainty due to the background shape is estimated by replacing the linear function with a second-order polynomial in the fit. The maximum difference in the signal yields compared to the nominal value is taken as the uncertainty.

    \item $\eta^{\prime}$ sideband. The uncertainties from the $\eta^{\prime}$ sideband region are estimated by changing sideband regions to [0.93, 0.94] and [0.975, 0.985] GeV/$c^{2}$. The maximum difference in signal yields compared to the nominal value is taken as the uncertainty.

    \item Peaking background.  The uncertainties from the peaking background are estimated by changing the fixed number of peaking background events, which is calculated through the propagation of branching fraction errors. The fixed branching fraction values of peaking background are calculated by adding or subtracting $1\sigma$ from the nominal value of these processes.
    The maximum difference in signal yields compared to the nominal value is taken as the uncertainty.
\end{enumerate}

For the two $\eta^{\prime}$ decay modes, some systematic uncertainties are common, while others are independent. The combination of common and independent systematic uncertainties for these two decay modes are calculated using the weighted least squares method \cite{DAgostini:1993arp}.

For the calculation of $\mathcal{B}(\chi_{c1,\, c2} \to\eta\eta\eta^{\prime})$, the sources of systematic uncertainties are listed in \text{Table \ref{tab_chic12}}.
Their branching fractions are determined to be 
$\mathcal{B}(\chi_{c1}\to\eta\eta\eta^{\prime})=(1.40\, \pm 0.13\, (\text{stat.}) \pm 0.09\, (\text{sys.}))\times 10^{-4}$ 
and 
$\mathcal{B}(\chi_{c2}\to\eta\eta\eta^{\prime})=(4.18\, \pm 0.84\, (\text{stat.})  \pm 0.48\, (\text{sys.}))\times 10^{-5}$, respectively.
It is noted that $\mathcal{B}(\chi_{c1}\to\pi^+\pi^-\eta^{\prime})=(2.2\pm 0.4)\times 10^{-3}$ and $\mathcal{B}(\chi_{c2}\to\pi^+\pi^-\eta^{\prime})=(5.1\pm 1.9)\times 10^{-4}$ \cite{ParticleDataGroup:2024cfk}. In comparison, the branching fractions of $\chi_{c1,\, c2}\to\eta\eta\eta^{\prime}$ obtained in this analysis are an order of magnitude smaller than those of $\chi_{c1,\, c2}\to\pi^+\pi^-\eta^{\prime}$.

\begin{table}[!htbp]
	\setlength{\abovecaptionskip}{0cm}
	\setlength{\belowcaptionskip}{0.5cm}
    \small
    \centering
    \caption{Summary of relative systematic uncertainties for $\mathcal{B}(\chi_{c1, c2}\to\eta\eta\eta^{\prime})$ (in \%). The symbol "-" denotes unapplicable items.}
    \label{tab_chic12}
\scalebox{0.8}{
\begin{tabular}{c|cc|cc}
        \hline
        \hline
        \multicolumn{5}{c}{\textbf{Independent systematic uncertainties}} \\ \hline
        \multirow{2}{*}{Source} &  \multicolumn{2}{c|}{$\chi_{c1}$}  & \multicolumn{2}{c}{$\chi_{c2}$} \\
        & $\gamma\pi^{+}\pi^{-}$ & $\eta\pi^{+}\pi^{-}$ & $\gamma\pi^{+}\pi^{-}$ & $\eta\pi^{+}\pi^{-}$  \\
        \hline
Another photon reconstruction & - & 0.5 & - & 0.5 \\
$\mathcal{B}(\eta\to\gamma\gamma)$  & - & 0.5 & - & 0.5 \\
$\mathcal{B}(\eta^{\prime}\to\gamma(\eta)\pi^{+}\pi^{-})$  & 1.4 & 1.2 & 1.4 & 1.2 \\
Kinematic fit   & 0.7 & 0.1 & 0.9 & 0.2 \\
$\eta^{\prime}$ mass window  & 0.1 & 0.1 & 0.1 & 0 \\
\hline
Total & 1.6 & 1.4 & 1.7 & 1.4 \\
\hline
        \multicolumn{5}{c}{\textbf{Common systematic uncertainties}} \\
        \hline
        Source & \multicolumn{2}{c|}{$\chi_{c1}$} & \multicolumn{2}{c}{$\chi_{c2}$} \\
        \hline
        $N_{\psi(3686)}$           &  \multicolumn{4}{c}{0.5} \\
        Two pion tracking          &  \multicolumn{4}{c}{2.0}  \\
        PID                        &  \multicolumn{4}{c}{2.0}  \\
        Six photon reconstruction  &  \multicolumn{4}{c}{3.0}   \\
        $\mathcal{B}(\eta\to\gamma\gamma)$    &  \multicolumn{4}{c}{1.0} \\
        \cline{2-5}
        $\mathcal{B}(\psi(3686)\to\gamma\chi_{cJ})$ & \multicolumn{2}{c|}{2.8} & \multicolumn{2}{c}{2.5} \\
Veto $\pi^{0}$ & \multicolumn{4}{c}{1.5} \\
Fit range           & \multicolumn{2}{c|}{1.4} & \multicolumn{2}{c}{4.1} \\
Background shape    & \multicolumn{2}{c|}{0.7} & \multicolumn{2}{c}{2.9} \\
$\eta^{\prime}$ sideband    & \multicolumn{2}{c|}{2.9} & \multicolumn{2}{c}{7.9} \\
Peaking background  & \multicolumn{2}{c|}{2.1} & \multicolumn{2}{c}{4.1} \\
\hline
Total &  \multicolumn{2}{c|}{6.6} & \multicolumn{2}{c}{11.5} \\
\hline
\textbf{Combined systematic uncertainties} & \multicolumn{2}{c|}{6.7} & \multicolumn{2}{c}{11.6} \\
    \hline
    \hline
\end{tabular}
}
\end{table}

The systematic uncertainties of $\mathcal{B}(\chi_{c0}\to\eta\eta\eta^{\prime})$ are divided into two categories.
The first category includes the multiplicative systematic uncertainties related to event selection, which are summarized in \text{Table \ref{tab_chic0}}.
The second category consists of the additive systematic uncertainties related to the fit, such as the fit range, background shape, $\eta^{\prime}$ sideband, and peaking background. 
To account for the additive systematic uncertainties related to the fits, several alternative fits are performed. Among the results of these fits, the largest number of signal events is chosen to calculate the upper limit.

\begin{table}[!htbp]
	\setlength{\abovecaptionskip}{0cm}
	\setlength{\belowcaptionskip}{0.5cm}
    \small
    \centering
    \caption{Summary of relative multiplicative systematic uncertainties for the upper limit of $\mathcal{B}(\chi_{c0}\to\eta\eta\eta^{\prime})$ (in \%).}
    \label{tab_chic0}

\begin{tabular}{c|cc}
        \hline
        \hline
        \multicolumn{3}{c}{\textbf{Independent multiplicative systematic uncertainties}} \\ \hline
        \multirow{2}{*}{Source} &  \multicolumn{2}{c}{$\chi_{c0}$} \\
        &$\gamma\pi^{+}\pi^{-}$ & $\eta\pi^{+}\pi^{-}$   \\
        \hline
Another photon reconstruction & - & 0.5 \\
$\mathcal{B}(\eta\to\gamma\gamma)$ & - & 0.5  \\
$\mathcal{B}(\eta^{\prime}\to\gamma(\eta)\pi^{+}\pi^{-})$ & 1.4 & 1.2 \\
Kinematic fit  & 1.0 & 0.2 \\
$\eta^{\prime}$ mass window & 0.1 & 0.1 \\
\hline
Total & 1.7 & 1.4\\
\hline
        \multicolumn{3}{c}{\textbf{Common multiplicative systematic uncertainties}} \\
        \hline
        Source & \multicolumn{2}{c}{$\chi_{c0}$}  \\
        \hline
        $N_{\psi(3686)}$           &  \multicolumn{2}{c}{0.5} \\
        Two pion tracking          &  \multicolumn{2}{c}{2.0}  \\
        PID                        &  \multicolumn{2}{c}{2.0}  \\
        Six photon reconstruction  &  \multicolumn{2}{c}{3.0}   \\
        $\mathcal{B}(\eta\to\gamma\gamma)$    &  \multicolumn{2}{c}{1.0} \\
        $\mathcal{B}(\psi(3686)\to\gamma\chi_{c0})$ & \multicolumn{2}{c}{2.4}  \\
        Veto $\pi^0$ & \multicolumn{2}{c}{1.5}   \\
\hline
Total & \multicolumn{2}{c}{5.1} \\
\hline
\textbf{Combined multiplicative uncertainties} & \multicolumn{2}{c}{5.2} \\
\hline
    \hline
    \end{tabular}
\end{table}

To obtain a conservative estimate of the upper limit of $\mathcal{B}(\chi_{c0}\to\eta\eta\eta^{\prime})$, the likelihood distribution is smeared by a Gaussian function with a mean of zero and a width of $\sigma_\epsilon$,
\begin{equation}
    \label{eq_Ln}
    L'(n')\varpropto \int_{0}^{1} L(n\frac{\epsilon}{\epsilon_0}) \rm exp [\frac{-(\epsilon-\epsilon_0)^2}{2\sigma_\epsilon^2}] d \epsilon ,
\end{equation}
where $L(n)$ is the likelihood distribution as a Gaussian function of the yield n, $\epsilon_0$ is the detection efficiency calculated by $\frac{\sum_{i}B_i \epsilon_i}{\sum_{i}B_i}$, and $\sigma_\epsilon$ is the multiplicative systematic uncertainty.
$L'(n')$ is the smeared likelihood distribution.
The upper limit of $\mathcal{B}(\chi_{c0}\to\eta\eta\eta^{\prime})$ is estimated to be $2.59\times 10^{-5}$.

\section{\boldmath Search for $\chi_{c1}\to\eta_1(1855)\eta, \eta_1(1855)\to\eta\eta^{\prime}$}

\begin{figure*}[!htbp]
    \centering
    \subfigure[]{\includegraphics[width=0.4\textwidth]{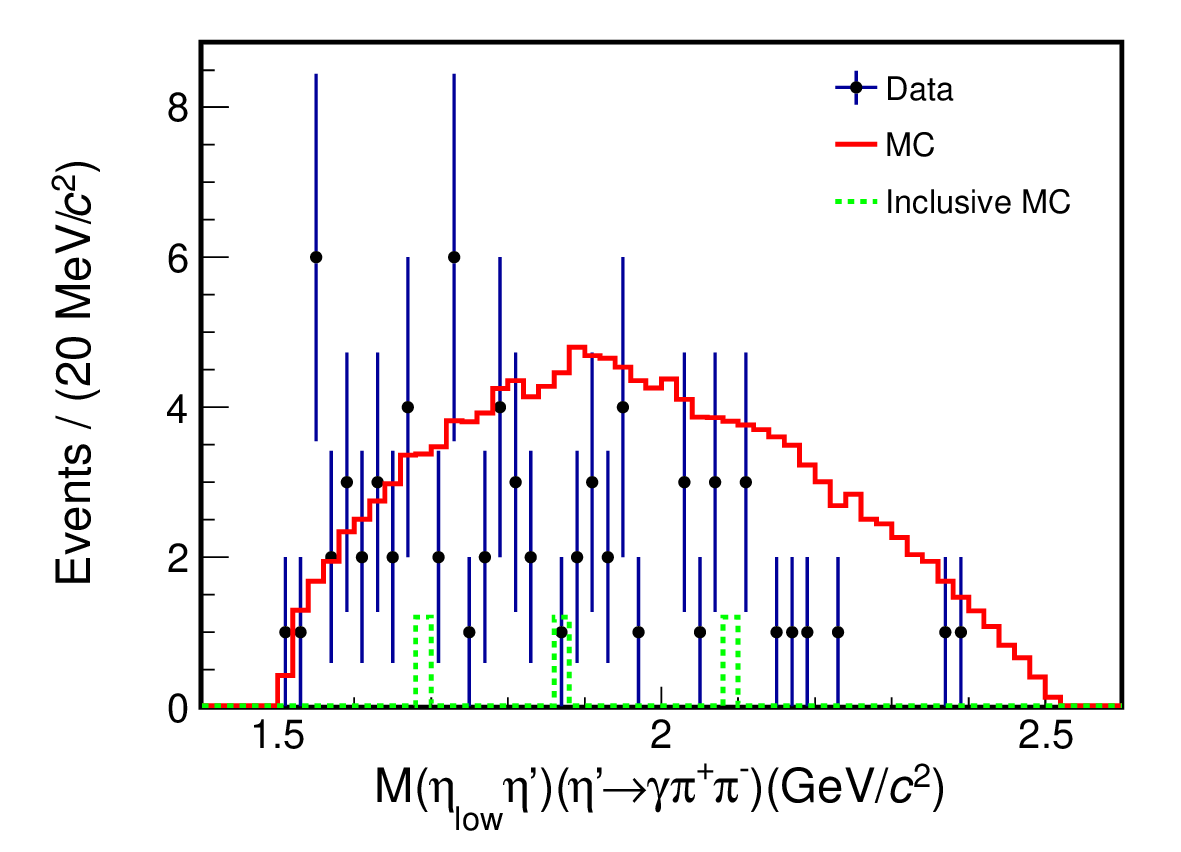}}
    \subfigure[]{\includegraphics[width=0.4\textwidth]{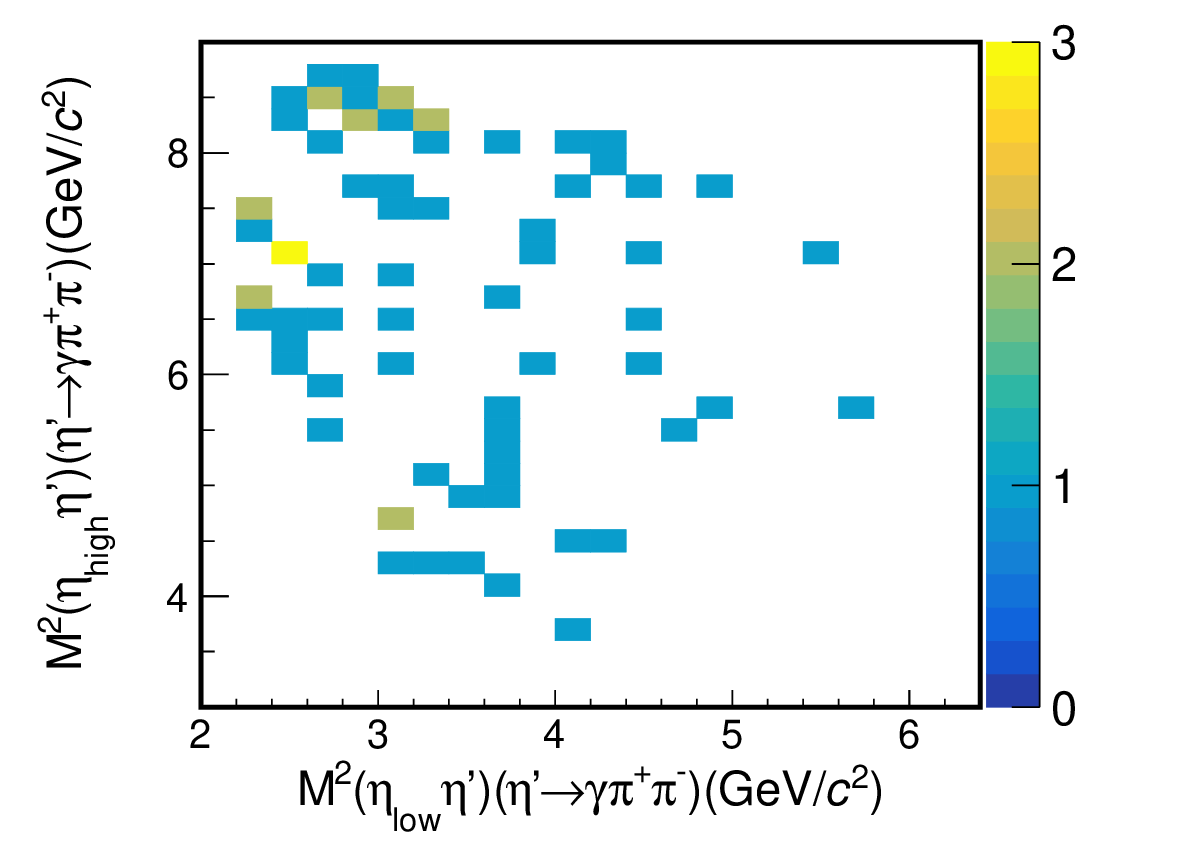}}   \\
    \subfigure[]{\includegraphics[width=0.4\textwidth]{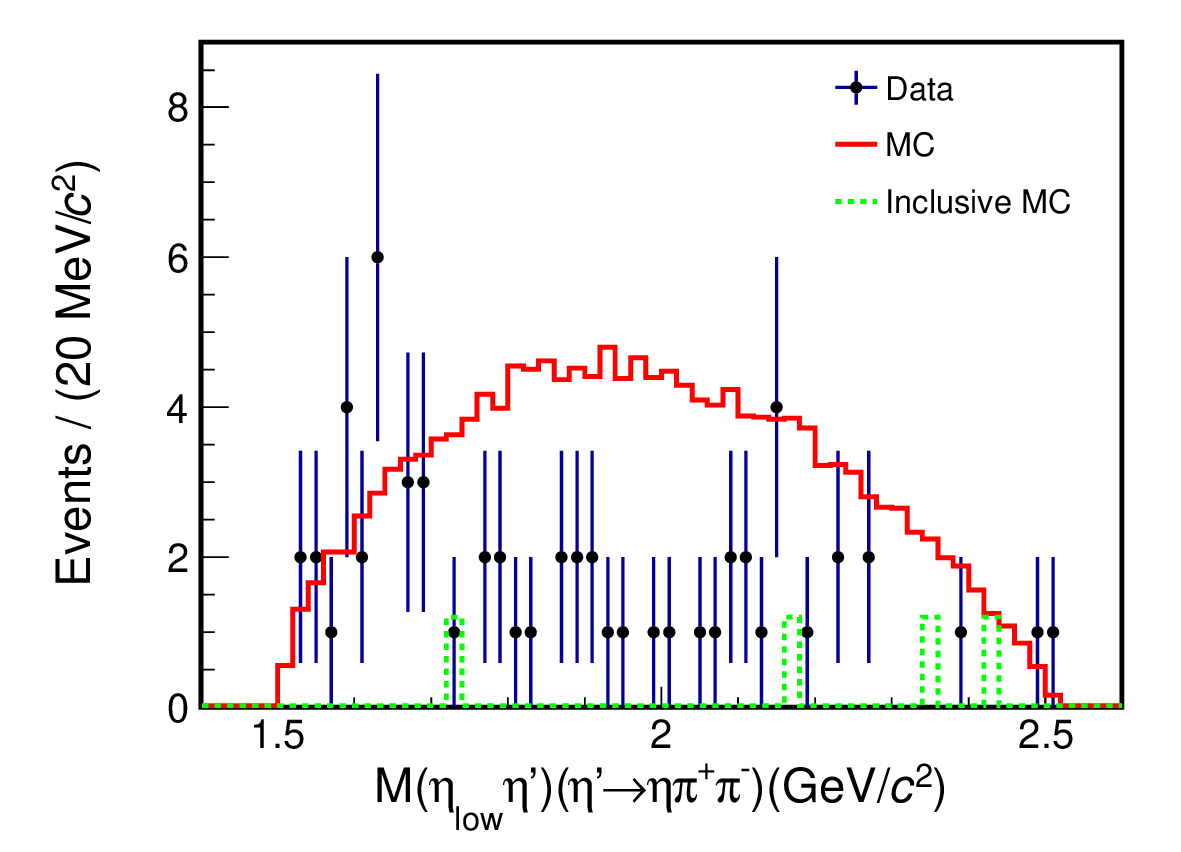}}
    \subfigure[]{\includegraphics[width=0.4\textwidth]{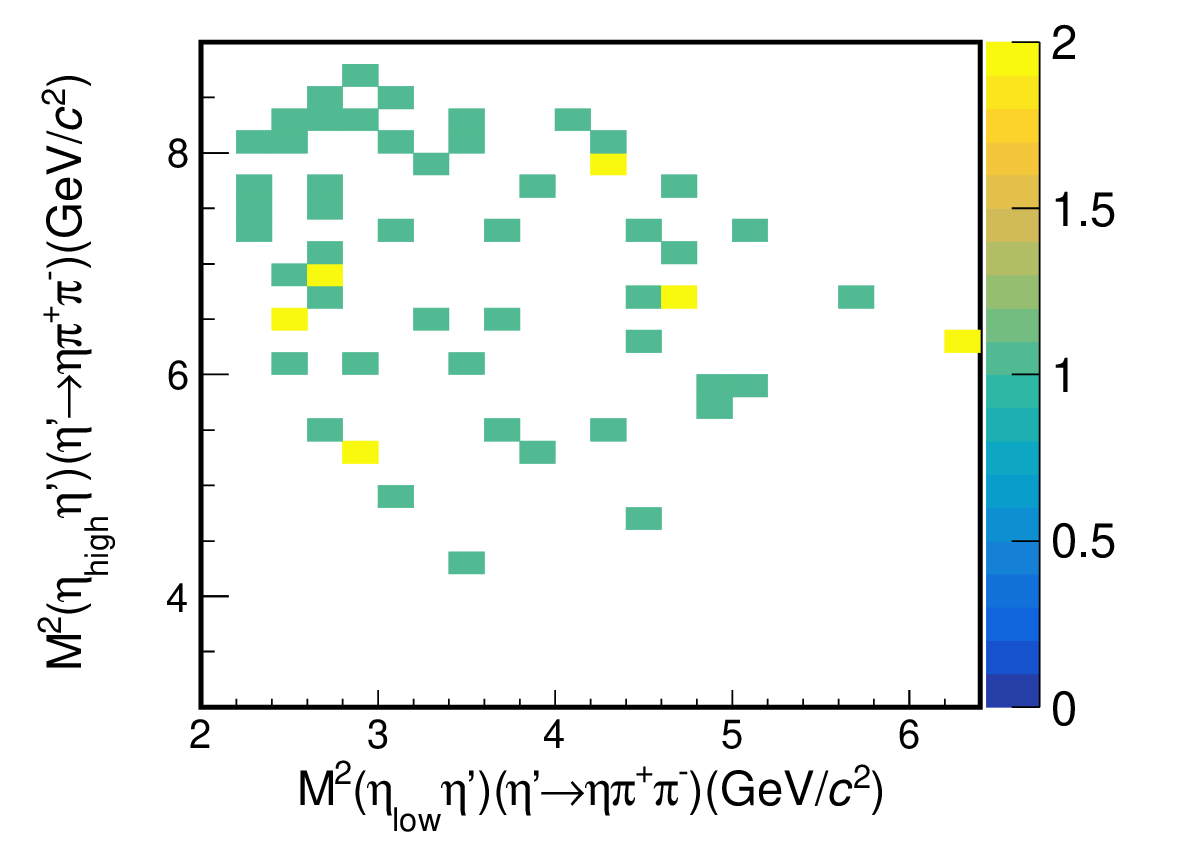}}   \\
    \caption{Invariant mass distributions and Dalitz plots for (a)(b) $\chi_{c1}\to\eta\eta\eta^{\prime}, \eta^{\prime}\to\gamma\pi^+\pi^-$, (c)(d)$\chi_{c1}\to\eta\eta\eta^{\prime}, \eta^{\prime}\to\eta\pi^+\pi^-$.
    The dots with error bars are data, the red solid histograms are the signal MC, and the green dashed histograms are inclusive MC. $\eta_{\text{low}}$ denotes the low energy $\eta$. }
    \label{fig_Mchic1plot}
\end{figure*}

To search for the $\chi_{c1}\to\eta_1(1855)\eta, \eta_1(1855)\to\eta\eta^{\prime}$ process, the $\eta\eta\eta^{\prime}$ invariant mass of the $\chi_{c1}$ candidates must distribute within  [3.50, 3.53] GeV/$c^{2}$. After the  event selections, the invariant mass spectra and Dalitz plots of data are shown in \text{Fig. \ref{fig_Mchic1plot}}.

\subsection{Background estimation}
Potential backgrounds are studied using the inclusive MC samples of $\psi(3686)$ decays.
The results show that the backgrounds can be divided into two categories: non-$\eta^{\prime}$ processes and those containing $\eta^{\prime}$.
Non-$\eta^{\prime}$ backgrounds are estimated using the $\eta^{\prime}$ mass sidebands in the data.
The sideband regions of $\eta^{\prime}$ are defined as [0.925, 0.935] and [0.98, 0.99] GeV/$c^{2}$.
The normalization factors for events in the two sideband regions are obtained by fitting the invariant mass distributions of $\gamma\pi^{+}\pi^{-}$ and $\eta\pi^{+}\pi^{-}$ in the data.
The $\eta^{\prime}$ signal shapes are determined from the shapes of signal MC samples, while the $\eta^{\prime}$ sideband shapes are described using linear functions.
The backgrounds containing $\eta^{\prime}$ are peaking backgrounds.
For this type of background, the branching fractions from the PDG and the efficiency obtained from background MC samples are used to estimate the normalized number of events, which is then fixed in the fit.
The peaking backgrounds for the decay mode $\eta^{\prime}\to\gamma\pi^+\pi^-$ mainly come from $\chi_{c1}\to\eta J/\psi, J/\psi\to\gamma\eta\eta^{\prime}$ and $\chi_{c1}\to\gamma J/\psi, J/\psi\to\gamma\eta\eta^{\prime}$, with estimated yields (fractions) of 0.2 (0.2\%) and 1.9 (2.6\%), respectively.
Similarly, the peaking backgrounds for the decay mode $\eta^{\prime}\to\eta\pi^+\pi^-$ mainly come from $\chi_{c1}\to\gamma J/\psi, J/\psi\to\gamma\eta\eta’$ and $\chi_{c0}\to\eta^{\prime}\eta^{\prime}$, with estimated yields (fractions) of 1.5 (2.6\%) and 0.2 (0.3\%), respectively.
The background with a fraction of no more than 1.0\% is neglected.

\subsection{PWA method}
For the two $\eta^{\prime}$ decay modes, there are 73 and 59 events left in the data, which are used to perform the PWA fit.
The four-momenta of $\psi(3686)$, $\gamma$, $\eta$, $\eta$ and $\eta^{\prime}$ after the 5C kinematic fit are used in the PWA.
The decay amplitude is constructed using the helicity amplitude formalism, and the full procedure is implemented based on the open-source framework TF-PWA\cite{Jiang:2024vbw}.
To construct the full decay amplitude of $\psi(3686)\to\gamma\eta\eta\eta^{\prime}$, the helicity formalism is used in conjunction with the isobar model, where the four-body decay is described as a two-step sequential quasi-two-body decay. For each two-body decay $A\to B+C$, the helicity amplitude can be written as:
\begin{equation}
    A^{A\to B+C}_{\lambda_A, \lambda_B, \lambda_C} = F_{\lambda_B \lambda_C} D^{j_{A^\ast}}_{\lambda_A, \lambda_{B-C}}(\phi, \theta, 0),
\end{equation}
where $D^{j_{A^\ast}}_{\lambda_A, \lambda_{B-C}}(\phi, \theta, 0)$ represents the Wigner D-function which describes the the angular distribution of the final state particle with its polar ($\theta$) and azimuthal ($\phi$) angles in the rest frame of the mother particle, and $F_{\lambda_B \lambda_C} $ is the helicity coupling amplitude, which can be described using the LS coupling formula,
\begin{equation}
    \begin{split}
        &F_{\lambda_B\lambda_C}=\sum_{ls}g_{ls}\sqrt{\frac{2l+1}{2J+1}}<j_B\lambda_B;j_c-\lambda_c|\mathrm{s}\delta> \\
        &<l0;\mathrm{s}\delta|J\delta>\left.q^lB_l^{\prime}(q,q_0,d)\right. ,
    \end{split}
\end{equation}
where $g_{ls}$ represents the fitting parameter, $\lambda_{A,B,C}$ is the helicity of particle A, B and C, $q$ and $q_0$ are the momentum of the invariant mass and the mass of the resonance state in the rest frame, and $B_l^{\prime}(q,q_0,d)$ is the Blatt-Weisskopf barrier factor.

The full amplitude for the complete decay chain $A\to R_1+B, R_1\to R_2+C, R_2\to D+E$ is constructed as the product of each two-body decay amplitude and the resonant propagator $R_1 (\chi_{c1})$ and $R_2$. The amplitude is written as Eq. \ref{eq_amplitude}, where $A,B,C,D,E$ are $\psi(3686),\gamma,\eta,\eta$ and $\eta^{\prime}$, respectively.
\begin{figure*}[ht]
    \centering
    \hrulefill
    \begin{equation}
        \label{eq_amplitude}
        \begin{split}
        & A_{\lambda_A,\lambda_B',\lambda_C',\lambda_D',\lambda_E'}^{R_1,R_2}
        = F_{{\lambda_{{R_{1}\lambda_{B}}}}}D_{{\lambda_{A},\lambda_{{R_{1}}}-\lambda_{B}}}^{{j_{{A^{*}}}}}(\boldsymbol{\varphi}_{1},\boldsymbol{\theta}_{1},\boldsymbol{0})R_{1}(\boldsymbol{M})
        F_{{\lambda_{{R_{2}}}\lambda_{C}}}D_{{\lambda_{{R_{1}}},\lambda_{{R_{2}}}-\lambda_{C}}}^{{j_{{R_{1}^{*}}}}}(\boldsymbol{\varphi}_{2},\boldsymbol{\theta}_{2},0)
        R_{2}(\boldsymbol{M}) F_{{\lambda_{D}\lambda_{E}}}D_{{\lambda_{{R_{2}}},\lambda_{D}-\lambda_{E}}}^{{j_{{R_{2}^{*}}}}}(\boldsymbol{\varphi}_{3},\boldsymbol{\theta}_{3},0) \\
        & D_{{\lambda_{B},\lambda_{B'}}}^{{j_{{B^{*}}}}}(\boldsymbol{\alpha}_{B},\boldsymbol{\beta}_{B},\boldsymbol{\gamma}_{B})D_{{\lambda_{C},\lambda_{C'}}}^{{j_{{C^{*}}}}}(\boldsymbol{\alpha}_{C},\boldsymbol{\beta}_{C'},\boldsymbol{\gamma}_{C})
        D_{{\lambda_{D},\lambda_{D'}}}^{{j_{{D^{*}}}}}(\boldsymbol{\alpha}_{D},\boldsymbol{\beta}_{D},\boldsymbol{\gamma}_{D})D_{{\lambda_{E},\lambda_{E'}}}^{{j_{{E^{*}}}}}(\boldsymbol{\alpha}_{E},\boldsymbol{\beta}_{E},\boldsymbol{\gamma}_{E})
        .
        \end{split}
    \end{equation}
\end{figure*}
The propagator R uses the relativistic Breit-Wigner formula with constant width: $BW(s)=\frac{1}{M^2 - s - iM\Gamma}$, where $M$ and $\Gamma$ are the mass and width of resonances, and $\sqrt{s}$ is the invariant mass of $\eta\eta^{\prime}$ or $\eta\eta$.
Backgrounds in the PWA are estimated by using normalized $\eta^{\prime}$  sidebands and the fixed MC shape in the processes of peaking backgrounds.

The construction of the probability density function, the calculation of the fit fraction, and the corresponding statistical uncertainty for each component follow Ref.\cite{Oset:2001cn}.
The combined branching fraction is obtained according to Eq. \ref{eq_TFPWAFraction}.

\begin{figure*}
    \centering
    \begin{equation}
        \label{eq_TFPWAFraction}
        \begin{aligned}&\mathcal{B}(\chi_{c1}\to X\eta(\eta^{\prime}))\cdot \mathcal{B}\big(X\to\eta\eta^{\prime}(\eta\eta)\big)\\&=\frac{N_{X_a}+N_{X_b}}{N_{\psi(3686)}\cdot \mathcal{B}(\psi(3686)\to\gamma\chi_{c1})\cdot \mathcal{B}^2(\eta\to\gamma\gamma)\cdot(\epsilon_{X_a}\cdot \mathcal{B}(\eta^{\prime}\to\gamma\pi\pi)+\epsilon_{X_b}\cdot \mathcal{B}(\eta^{\prime}\to\eta\pi\pi))\cdot \mathcal{B}\left(\eta\to\gamma\gamma\right))}.\end{aligned}
    \end{equation}
\end{figure*}

\subsection{PWA result}
From the mass spectrum, no obvious structure is observed in the $\eta\eta$ mass spectrum, while a structure is seen near the threshold in the mass spectrum of $\eta\eta^{\prime}$.
Therefore, the S-wave nonresonant (NR) component and $f_0(1500)$ in the $\eta\eta$ mass spectrum, as well as the S-wave NR component, $f_0(1500)$, and $\eta_1(1855)$ in the $\eta\eta^{\prime}$ mass spectrum, are considered.
The mass and width of $\eta_1(1855)$ are fixed at $1855$ MeV/$c^2$ and 188 MeV/$c^2$ as given in the Ref. \cite{BESIII:2022riz,BESIII:2022iwi}.
The statistical significance for each component is determined by examining the change in negative log-likelihood values when this resonance is included or excluded in the fits. The probability is calculated under the $\chi^2$ distribution hypothesis considering the change in the number of degrees of freedom.
Resonance with a statistical significance greater than $5\sigma$ is included in the final set of amplitudes.
Since the widths of the resonances in the PWA are sufficiently broad, the effect of detector resolution is neglected in the partial wave analysis.

\begin{figure*}[!htbp]
    \centering
    \subfigure[]{\includegraphics[width=0.32\textwidth]{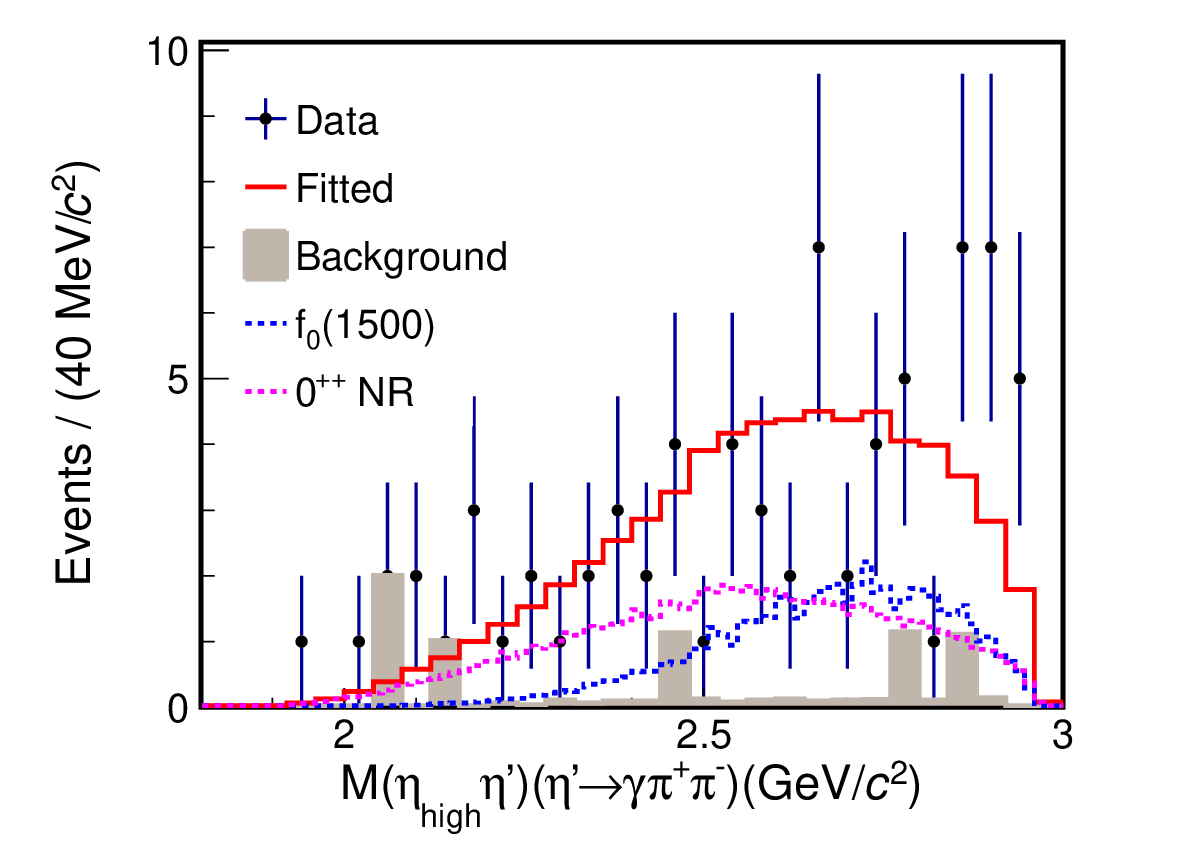}}
    \subfigure[]{\includegraphics[width=0.32\textwidth]{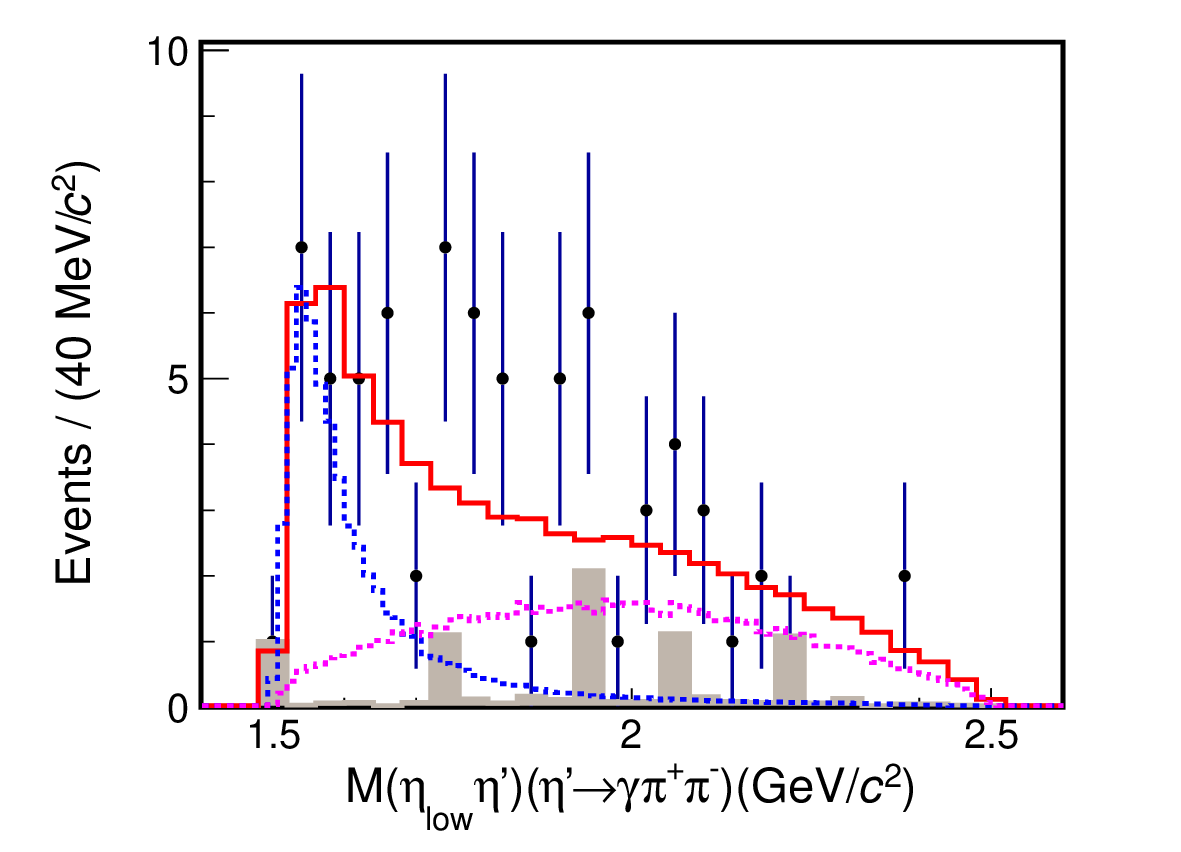}}
    \subfigure[]{\includegraphics[width=0.32\textwidth]{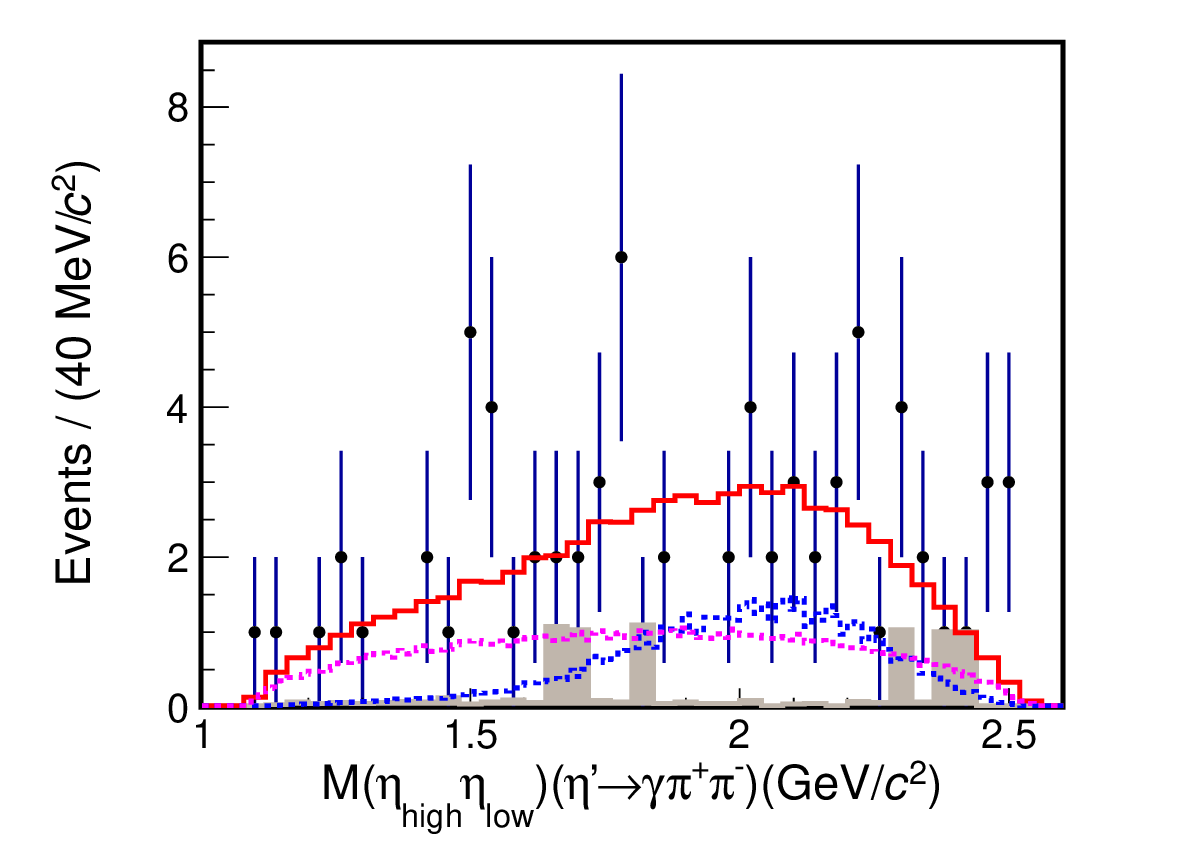}}
    \subfigure[]{\includegraphics[width=0.32\textwidth]{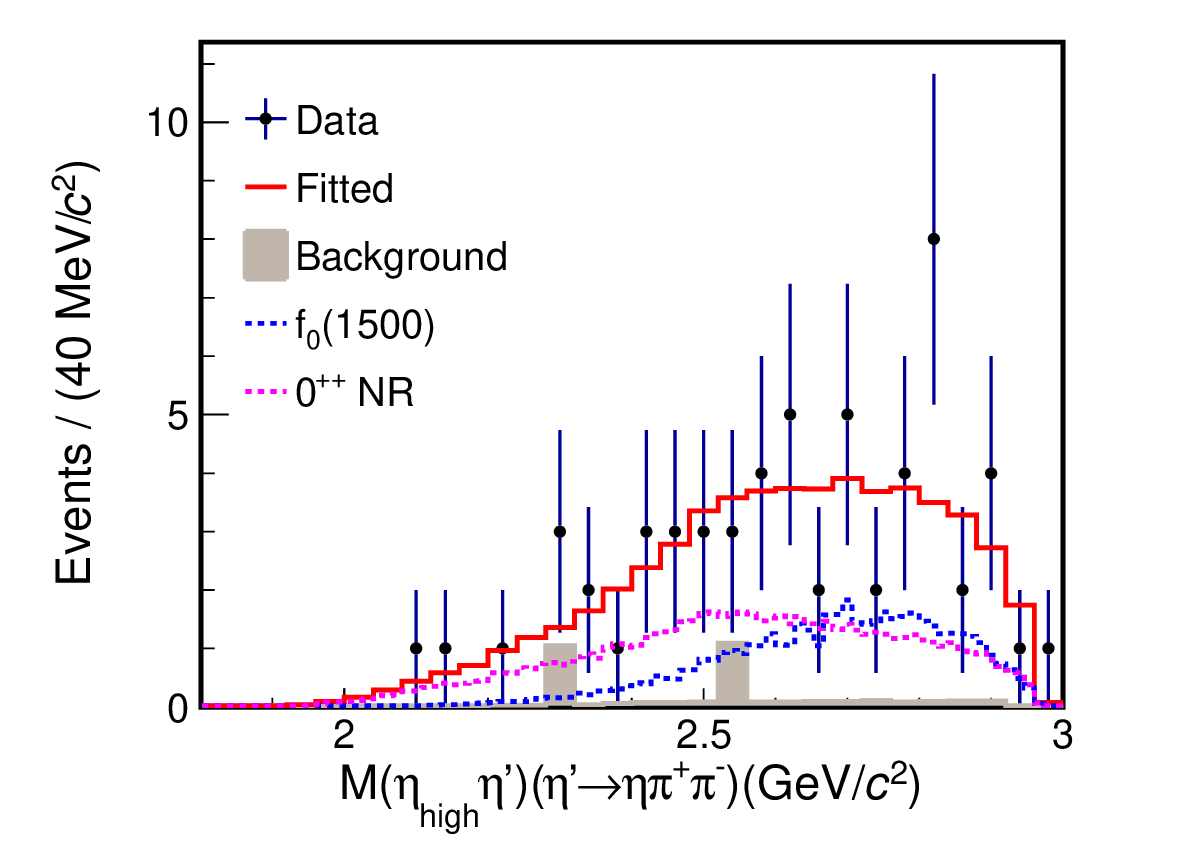}}
    \subfigure[]{\includegraphics[width=0.32\textwidth]{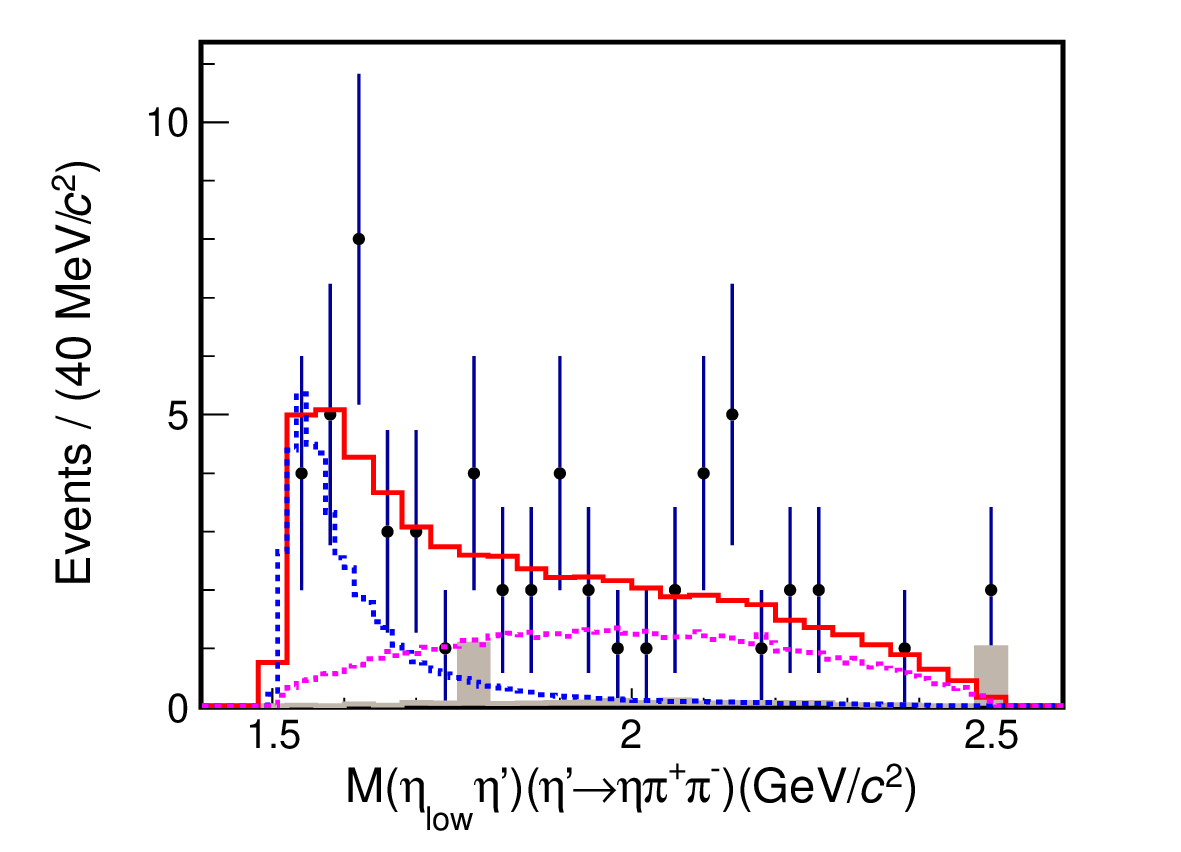}}
    \subfigure[]{\includegraphics[width=0.32\textwidth]{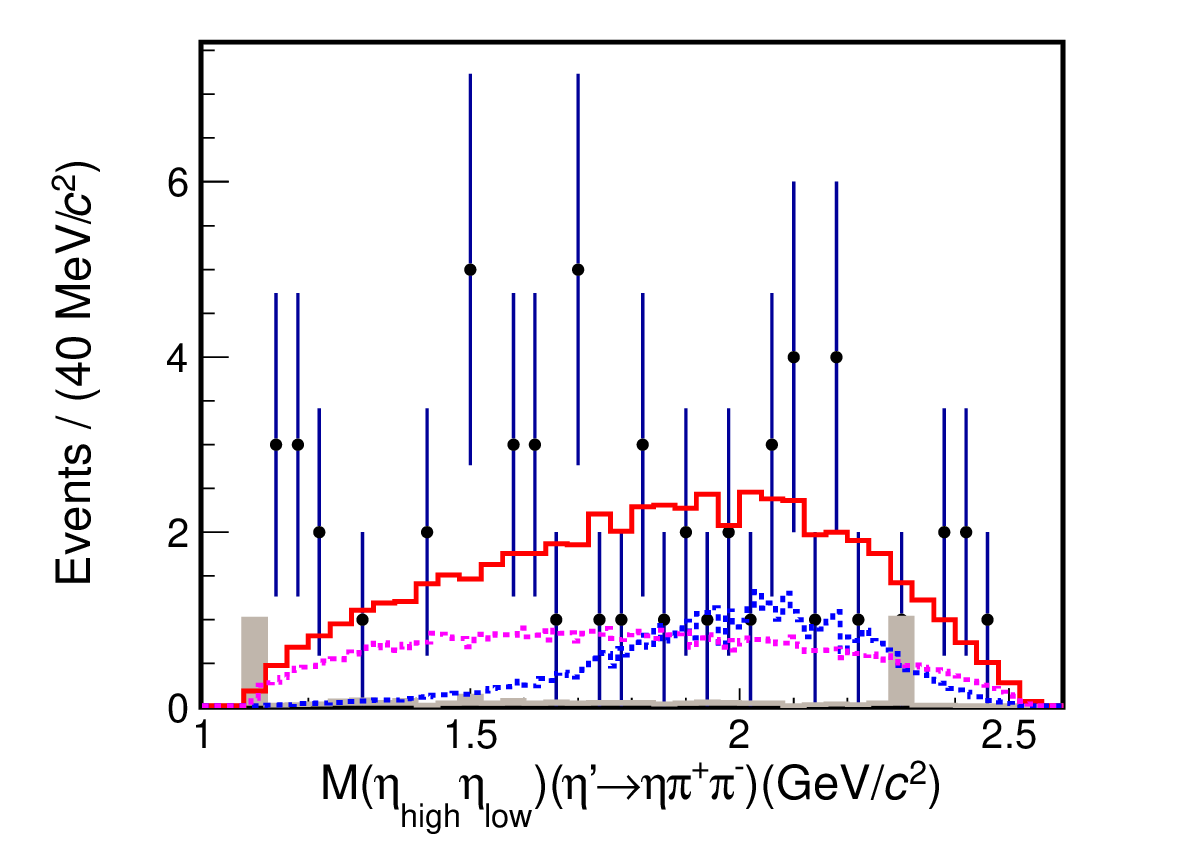}}
    \caption{Data (black points with error bars) and the PWA fit projections (lines) for (a)(b)(c) $\chi_{c1}\to\eta\eta\eta^{\prime}, \eta^{\prime}\to\gamma\pi^+\pi^-$ and (d)(e)(f) $\chi_{c1}\to\eta\eta\eta^{\prime}, \eta^{\prime}\to\eta\pi^+\pi^-$. The invariant mass distributions of (a)(d) $\eta_{\text{high}}\eta^{\prime}$, (b)(e) $\eta_{\text{low}}\eta^{\prime}$ and (c)(f) $\eta_{\text{high}}\eta_{\text{low}}$. The red solid lines are the total fit projections from the PWA, the gray shadows are the background (from $\eta^{\prime}$ sideband and peaking background), the blue dashed lines are the $f_0(1500)$ and the pink dashed lines are the $0^{++}$ NR components.}
    \label{fig_tfpwa1}
\end{figure*}

The results of the PWA show that the structure in the $\eta\eta^{\prime}$ mass spectrum is dominated by $f_0(1500)$ with the statistical significance of $6.5\sigma$, while the structure of $\eta\eta$ is mainly described by the $0^{++}$ NR component.
\text{Fig. \ref{fig_tfpwa1}} displays the projection plots of the fitting results of the PWA basic solution.
The statistical significance of an additional $\eta_1(1855)$ resonance is determined to be $0.7\sigma$.
An upper limit of branching fraction $\mathcal{B}(\chi_{c1}\to\eta_{1}(1855)\eta) \cdot \mathcal{B}(\eta_{1}(1855)\to \eta\eta^{\prime})$ is set. 
The systematic uncertainties will be described below.

\subsection{Systematic uncertainties}
The systematic uncertainties are divided into two categories.
The first category consists of the multiplicative systematic uncertainties related to event selection, summarized in \text{Table \ref{tab_systfpwa1}}.
The second category arises from the PWA, with each alternative fit discussed below.

\begin{enumerate}
    \item Uncertainties from $\eta^{\prime}$ sideband background and peaking background are treated as before.
    \item Uncertainty from additional resonances.
    To investigate the impact of other possible components on the PWA results, $0^{++}$ NR component is added to the mass spectrum of $\eta\eta^{\prime}$, or $f_0(1500)$ is added to the mass spectrum of $\eta\eta$.
    \item Uncertainty from resonance parameters.
    In the baseline fit, the resonance parameters of the $\eta_1(1855)$ are fixed to PDG values \cite{ParticleDataGroup:2024cfk}. An alternative fit is performed where resonance parameters are allowed to vary within one standard deviation of the PDG values, and the changes in the results are taken as systematic uncertainties.
\end{enumerate}

\begin{table}[H]
	\setlength{\abovecaptionskip}{0cm}
	\setlength{\belowcaptionskip}{0.5cm}
    \small
    \centering
    \caption{Summary of relative multiplicative systematic uncertainties for $\mathcal{B}(\chi_{c1}\to\eta\eta_1(1855),\eta_1(1855)\to\eta\eta^{\prime})$ (in \%).}
    \label{tab_systfpwa1}
\begin{tabular}{c|cc}
        \hline
        \hline
        \multicolumn{3}{c}{\textbf{Independent multiplicative systematic uncertainties}} \\
        \hline
        Source &$\gamma\pi^{+}\pi^{-}$ & $\eta\pi^{+}\pi^{-}$   \\
        \hline
Another photon reconstruction & - & 0.5 \\
$\mathcal{B}(\eta\to\gamma\gamma)$ & - & 0.5  \\
$\mathcal{B}(\eta^{\prime}\to\gamma(\eta)\pi^{+}\pi^{-})$ & 1.4 & 1.2 \\
Kinematic fit  & 0.8 & 0.1 \\
$\eta^{\prime}$ mass window & 0.1 & 0 \\
$\chi_{c1}$ mass window & 0.1 & 0 \\
\hline
Total & 1.6 & 1.4\\
\hline
        \multicolumn{3}{c}{\textbf{Common multiplicative systematic uncertainties}} \\
        \hline
        Source &  $\gamma\pi^{+}\pi^{-}$ & $\eta\pi^{+}\pi^{-}$   \\
        \hline
        $N_{\psi(3686)}$           &  \multicolumn{2}{c}{0.5} \\
        Two pion tracking          &  \multicolumn{2}{c}{2.0}  \\
        PID                        &  \multicolumn{2}{c}{2.0}  \\
        Six photon reconstruction  &  \multicolumn{2}{c}{3.0}   \\
        $\mathcal{B}(\eta\to\gamma\gamma)$    &  \multicolumn{2}{c}{1.0} \\
        $\mathcal{B}(\psi(3686)\to\gamma\chi_{c1})$ & \multicolumn{2}{c}{2.8}  \\
        Veto $\pi^0$    & \multicolumn{2}{c}{1.5}   \\
\hline
Total & \multicolumn{2}{c}{5.3} \\
\hline
        \textbf{Combined multiplicative uncertainties} & \multicolumn{2}{c}{5.4} \\
    \hline
    \hline
    \end{tabular}
\end{table}

For the systematic uncertainties associated with the PWA, the maximum value of the upper limit is chosen for each case.
Considering multiplicative systematic uncertainties associated with event selections, the likelihood distribution is smeared using a Gaussian function with a mean of zero and a width of $\sigma_{\epsilon}$, as Eq. \ref{eq_Ln},
where $L(n)$ is a Gaussian distribution with mean and error from the PWA branching fraction, $\epsilon_0$ is the detection efficiency obtained by PWA, and $\sigma_{\epsilon}$ is the combined multiplicative systematic uncertainty.
$L'(n')$ is the smeared likelihood distribution.
Taking the smearing distribution with an area of 90\% as the upper limit, the result is $\mathcal{B}(\chi_{c1}\to\eta_{1}(1855)\eta) \cdot \mathcal{B}(\eta_{1}(1855)\to\eta\eta^{\prime})<9.79\times 10^{-5}$.

\section{\boldmath Summary}
In summary, a study of the decays $\chi_{cJ}\to\eta\eta\eta^{\prime}$ and a PWA of $\chi_{c1}\to\eta\eta\eta^{\prime}$ are performed based on $2.7\times 10^9$ $\psi(3686)$ events collected with the BESIII detector.
The decay modes $\chi_{c1,\, c2}\to\eta\eta\eta^{\prime}$ are observed for the first time, and their corresponding branching fractions are determined to be $\mathcal{B}(\chi_{c1}\to\eta\eta\eta^{\prime}) = (1.40\,  \pm 0.13\, (\text{stat.}) \pm 0.09\, (\text{sys.})) \times 10^{-4}$ and 
$\mathcal{B}(\chi_{c2}\to\eta\eta\eta^{\prime}) = (4.18\,  \pm 0.84\, (\text{stat.}) \pm 0.48\, (\text{sys.})) \times 10^{-5}$, which are one order of magnitude lower than those of $\chi_{c1,\, c2}\to\pi^+\pi^-\eta^{\prime}$ in PDG \cite{ParticleDataGroup:2024cfk}.
An upper limit of $2.59 \times 10^{-5}$ at 90\% CL is set for the decay $\chi_{c0}\to\eta\eta\eta^{\prime}$.
Subsequently, a PWA of the decay $\chi_{c1}\to\eta\eta\eta^{\prime}$ is performed to search for the $1^{-+}$ exotic state $\eta_1(1855)$.
The PWA result indicates that the structure in the $\eta\eta^{\prime}$ mass spectrum is mainly attributed to $f_0(1500)$, while in the $\eta\eta$ mass spectrum, it is primarily the $0^{++}$ NR component.
No significant $\eta_1(1855)$ is observed.
The upper limit of $\mathcal{B}(\chi_{c1}\to\eta_{1}(1855)\eta) \cdot \mathcal{B}(\eta_{1}(1855)\to\eta\eta^{\prime})< 9.79 \times 10^{-5}$ is obtained at 90\% CL.
In further research, the $\eta\eta^{\prime}$ decay mode with an increased dataset and more decay modes will be explored to study $\eta_1(1855)$.

\section{\boldmath Acknowledgment}

The BESIII Collaboration thanks the staff of BEPCII (https://cstr.cn/31109.02.BEPC) and the IHEP computing center for their strong support. This work is supported in part by National Key R\&D Program of China under Contracts Nos. 2023YFA1606000, 2023YFA1606704; National Natural Science Foundation of China (NSFC) under Contracts Nos. 11635010, 11935015, 11935016, 11935018, 12025502, 12035009, 12035013, 12061131003, 12192260, 12192261, 12192262, 12192263, 12192264, 12192265, 12221005, 12225509, 12235017, 12361141819; the Chinese Academy of Sciences (CAS) Large-Scale Scientific Facility Program; CAS under Contract No. YSBR-101; 100 Talents Program of CAS; The Institute of Nuclear and Particle Physics (INPAC) and Shanghai Key Laboratory for Particle Physics and Cosmology; Agencia Nacional de Investigación y Desarrollo de Chile (ANID), Chile under Contract No. ANID PIA/APOYO AFB230003; German Research Foundation DFG under Contract No. FOR5327; Istituto Nazionale di Fisica Nucleare, Italy; Knut and Alice Wallenberg Foundation under Contracts Nos. 2021.0174, 2021.0299; Ministry of Development of Turkey under Contract No. DPT2006K-120470; National Research Foundation of Korea under Contract No. NRF-2022R1A2C1092335; National Science and Technology fund of Mongolia; National Science Research and Innovation Fund (NSRF) via the Program Management Unit for Human Resources \& Institutional Development, Research and Innovation of Thailand under Contract No. B50G670107; Polish National Science Centre under Contract No. 2024/53/B/ST2/00975; Swedish Research Council under Contract No. 2019.04595; U. S. Department of Energy under Contract No. DE-FG02-05ER41374.

\bibliographystyle{apsrev4-2}
\bibliography{reference}

\end{document}